\documentstyle{article}

\begin{document}

ON THE GENERAL PROPERTIES OF MATTER, by M\'{a}rio Everaldo de Souza,
Departamento de F\'{\i}sica, Universidade Federal de Sergipe, Campus
Universit\'{a}rio, 49000 Aracaju, Sergipe, Brazil, e-mail DFIMES@BRUFSE.BITNET

\vskip .25in

\noindent
I. INTRODUCTION
\vskip .15in
\par It has been proven beyond any doubt that the universe is expanding
$^{(1,2,3,4)}$. Recent data of several investigators show that  galaxies form
gigantic structures in space. De Lapparent et al.
$^{(5)}$(also, other papers by
the same authors) have shown that they form bubbles which contain huge voids of
many megaparsecs of diameter. Broadhurst et al.$^{(6)}$
probed deeper regions of the universe with two pencil
beams and showed that there are (bubble) walls up to a distance of about 2.5
billion light-years from our galaxy.  Even more disturbing is the apparent
regularity of the walls with a period of about
$130h^{-1}$Mpc. There is also an
agglomeration of galaxies forming a thicker wall, called the great wall. It has
also been reported$^{(7,8,9)}$ (also, other confirming papers) that there is a
large-scale coherent flow towards a region named the Great Attractor. All this
data show that  galaxies form a medium in which we already observe some
interactions.  Recent data$^{(10)}$  show, however, that the bubble walls are
not so regularly spaced and, therefore, the medium formed by them is rather a
liquid(or an amorphous solid) than a crystalline solid. We may call this medium
the `galactic liquid'.
\par The facts shown above have also been corroborated by the data of the APM
survey$^{(11)}$ which was based on a sample of more than two million galaxies.
This survey measured the galaxy correlation function $w(\theta)$.
More evidence towards the same conclusion was provided by the `counts in cells'
of the QDOT survey$^{(12)}$ and more recently by the redshift(APM)
survey$^{(13)}$ and by the power spectrum inferred from the CfA survey$^{(14)}$
and from that inferred from the Southern Sky Redshift Survey$^{(15)}$.
The  data from these surveys show that galaxies are not randomly distributed in
space on large scales and put a shadow over the cold dark matter
model$^{(16,17)}$. There have been attempts to keep the cold dark matter
hypothesis but only at the cost of many {\it{ad hoc}} adjustments.  There are
even proposals of having more than one type of dark matter$^{(18)}$.
\par Another puzzle of Nature is the existence of strange objects, such as
quasars, BL Lacertae and Seyfert galaxies, and the fact of having spiral
galaxies with their mysterial arms. The evolution of galaxies does not fit
either in the general theoretical framework.
\par At the other end of the distance scale, in the fermi region, it appears
now that the quark is not elementary after all. This can be implied just
from their number which, now, stands at 18. Particle physics theorists
have already begun making models addressing this compositeness$^{(19)}$.
\vskip .25in
\noindent
II. GENERAL CLASSIFICATION OF MATTER
\vskip .2in
\large
\hskip 4in{\it{Ad unguem}}
\normalsize
\vskip .15in
\par Science has utilized specific empirical classifications of matter
which have revealed hidden laws and symmetries.  Two of the most known
classifications are the Periodic Table of the Elements and Gell-Mann's
classification of particles(which paved the way towards the quark model).
\par Let us go on the footsteps of Mendeleev and let us attempt to achieve a
general classification of matter, including all kinds of matter formed along
the universal expansion, and by doing so we may find the links between the
elementary particles and the large bodies of the universe.
\par It is well known that the different kinds of matter of nature appeared
at different epochs of the universal expansion, and that, they are imprints of
the different sizes of the universe along the expansion.
Taking a closer look at the different kinds of matter  we may classify them
as belonging to two different general states. One state is characterized by a
single unity with  angular momentum, and we may call it, the single state.
The angular momentum may either be the intrinsic angular momentum, spin, or the
orbital angular momentum.
The other state is characterized by some degree of correlation among the
interacting particles and may be called
the `polarized' state. The angular momentum may(or may not) be present in this
state.  In the single states we find the fundamental unities of matter that
make the polarized states. The different kinds of fundamental matter are the
building blocks of everything, {\it{stepwise}}.
In what follows we will not talk about the weak force since it
does not form any stable matter and is rather related to instability in matter.
As discussed in this paper it appears that along the universal expansion
nature made different building blocks which filled the space.  The weak force
did not form any building block and is out of the initial discussion.
As is well known this force is special in many other ways.  For example, it
violates parity in many decays and it has no ``effective potential''(or static
potential) as the other interactions do.  Besides, the weak force is known
to be left-handed, that is, particles experience this force only when their
spin direction is anti-aligned with their momentum.  Right-handed particles
appear to experience no weak interaction, although, if they have electric
charge, they may still interact electromagnetically. Later on we will include
the weak force into the discussion. The single state is made by only one kind
of fundamental force. In the polarized state one always finds two types of
fundamental forces, i.e., this state is a link between two single states. Due
to the interactions among the bodies(belonging to a particular single state)
one  expects other kinds of forces in the polarized state. In this fashion
we can form a chain from the quarks to the galactic superstructures and
extrapolate at the two ends towards the constituents of quarks and towards the
whole universe.
\par The kinds of matter belonging to the single states are the nucleons,
 the atom, the galaxies, etc. The `et cetera' will become clearer later on
in this article. In the polarized state one finds the quarks, the nuclei, the
gasses, liquids and solids, and  the galactic liquid.
 Let us, for example, examine the sequence nucleon-nucleus-atom. A
nucleon is made out of quarks and held together by means of the strong
force. The atom is made out of the nucleus and the electron(we will talk about
the electron later), and is held together by means of the
electromagnetic force. The nucleus, which is in the middle of the sequence, is
held together by the strong force(attraction among nucleons) and by the
electromagnetic force(repulsion among protons). In other
words, we may say that the nucleus is a compromise between these two forces.
Let us, now, turn to the sequence atom-(gas,liquid,solid)-galaxy. The gasses,
liquids and solids are also formed by two forces, namely, the electromagnetic
and the gravitational forces. Because the gravitational force is $10 {39}$
weaker than the electromagnetic force the polarization in gasses, liquids and
solids is achieved by the sole action of the electromagnetic force because
it has two signs.  But it is well known that gasses, liquids and solids are
unstable configurations of matter in the absence of gravity.  Therefore, they
are formed by the electromagnetic and gravitational forces.  The clumping
of hydrogen {\it{gas}} at some time in the history of the universe gave
origin to galaxies which are the biggest individual clumps of creation.
We arrive again at a single fundamental force that holds a galaxy together,
which is the gravitational force. There is always the same pattern: one goes
from one fundamental force which exists in a single unity(nucleon, atom,
galaxy) to two fundamental forces which coexist in a medium. The
interactions in the medium form a new unity in which the action of another
fundamental force appears. We are not talking any more about the previous unity
which exists inside the new unity(such as the nucleons in the nucleus of an
atom).
\par By placing all kinds of matter together in a table in the order of the
{\it{universal expansion}}  we can construct the two tables below,
one for the states and another for the fundamental forces.
\par In order to make the atom we need the electron besides the nucleus.
Therefore, just the clumping of nucleons is not enough in this case.
Let us just borrow the electron for now.  Therefore, it looks like that the
electron belongs to a separate class and is an elementary particle.
The above considerations may be summarized by the following: {\it{the different
kinds of building blocks of the Universe(at different times of the expansion)
are intimately related to the idea of filling space.}}  That is, depending on
the size of the Universe, it is filled with different unities.
\vskip .25in
\noindent
III. THE NUMBER OF FUNDAMENTAL FORCES OF NATURE
\vskip .15in
\par In order to keep the same pattern, which should be related to an
underlying symmetry, the tables reveal that there should be another force,
other  than the strong force, holding the quarks together, and that this force
alone should hold together the prequarks. Let us name it the superstrong force.
Also, for the `galactic liquid', there must be another fundamental force at
play. Because it must be much weaker than the gravitational force(otherwise,
it would already have been found on Earth) we expect it to be a very weak
force.  Let us call it the superweak force.
\par Summing up all fundamental forces  we arrive at {\it{six forces for
nature: the superstrong, the strong, the electromagnetic, the gravitational,
the superweak and the weak  forces}}.  We will see later on, at the end of the
article, that they are interrelated.
\vskip .25in
\noindent
IV. POLARIZATION IN MATTER
\vskip .15in
\par The polarized state is made by opposing forces, i.e., they represent a
compromise between an attractive force and a repulsive force.  Ordinary
matter(gasses, liquids and solids) is formed by the polarization of the
electromagnetic force.  Polarization is also present in nuclear matter which
may be described either in terms of the Seyler-Blanchard interaction or
according to the Skirme interaction. Both  give a
type of Van der Waals equation of state$^{(20,21)}$.  In order to keep the same
pattern we should expect to have a sort of compromise between the superstrong
force and the strong force. This compromise forms the quark. The polarization
in this case may be achieved by means of the exchange of a particle, which
may be the boson of the superstrong interaction.  Actually, it is shown
later on that this interaction is mediated by three bosons.
\par In order to have the `galactic liquid' it is also necessary to have
some sort of polarization.  This means that we need dipoles and because the
gravitational force is always attractive(and thus, can not be the source of
such dipoles), the superweak force must be repulsive during the universal
expansion.  This is consistent with the idea of the expansion itself.  That is,
the universal expansion must be caused by this repulsive force.
\par The bodies which form any polarized state exhibit some degree of
correlation among them. This degree of correlation is shown by the correlation
function which, in turn, is related to the interacting net potential among the
particles.  The potential has three general features: i) it has a
minimum which is related to the mean equilibrium positions of the interacting
particles; ii) it tends to zero as the separation among the particles
tends to infinity; and iii) it becomes repulsive at close distances.
The potential of the `galactic  liquid' must also have the
same general features.  Therefore, it is very important to determine the mean
equilibrium position of the galactic superstructures.
\par As is well known the general motion of the particles of a fluid is quite
complex and that is exactly what we are dealing with in the case of the
galactic superstructures.  It is worth reminding that up to now there is no
acceptable theory which describes the liquid state of ordinary matter and there
is no general theory which describes nuclear matter either.
\vskip .25in
\noindent
V. THE SUPERWEAK FORCE
\vskip .15in
\par Let us now try to find a possible mathematical expression for the
superweak force. There have been reports of a fifth force inferred from the
reanalysis of the E\"{o}tv\"{o}s experiment and from the mine-gravity
data$^{(22)}$.  The discrepancies suggest the existence of a composition
dependent intermediate-range force.
\par The potential energy of such hypothetical force is usually represented by
a Yukawa potential which, when added to the standard Newtonian potential
energy, becomes$^{(22)}$
\begin{eqnarray} \displaystyle V(r)& = &- \frac{Gm_{1}m_{2}}{r}\left(1 +
\alpha\exp{(-r/\lambda)}\right),\end{eqnarray}

\noindent
where $\alpha$ is the new coupling in units of gravity and $\lambda$ is its
range.  The dependence on composition can be made explicit by writing
${\alpha}=q_{i}q_{j}\zeta$ with
\begin{eqnarray} \displaystyle q_{i}& =& cos{\theta}(N+Z)_{i}/\mu_{i} +
sin{\theta}(N-Z)_{i}/\mu_{i},\end{eqnarray}

\noindent
where the new effective charge has been written as a linear combination of the
baryon number and nuclear isospin per atomic mass unit, and $\zeta$ is the
coupling constant in terms of $G$.
\par Until now most experimental results have not confirmed the existence of
this force$^{(23)}$, although they do not rule it out because its coupling
constant(s) may be smaller than previously thought. Later on we will propose an
estimation of the order of magnitude of the coupling constant of this
interaction. Of course, this force may only exist if there is a violation in
the weak equivalence principle, which has been proven to hold$^{(23)}$ to a
precision of one part in $10^{12}$. But it may be violated if the precision
goes just one more order of magnitude.  There are some experiments that will be
carried out within the next few years and that intend to test the equivalence
principle to a higher accuracy$^{(24)}$.  They may, then, reveal the existence
of the fifth force.
\par The superweak force proposed in this paper has the same character as the
one of the fifth force, but has an infinite range.  This means that the mass of
the mediating boson is zero. From the above expression for the fifth force
potential we may express the potential of the superweak force in terms of
the baryon numbers and isospins of two bodies i and j as
\begin{eqnarray}
 V(r,N,Z)& = & \left(A_{B}(N+Z)_{i}(N+Z)_{j} + A_{I}(N-Z)_{i}(N-Z)_{j} +
\right.
\nonumber \\
         &   & \left.+ A_{IB}((N+Z)_{i}(N-Z)_{j} +
(N+Z)_{j}(N-Z)_{i})\right)g^{2}\frac{\exp{(-r/\lambda)}}{r}\end{eqnarray}

\noindent
where $A_{B}$ and $A_{I}$ are the force coupling constants of the baryon number
and isospin  terms, respectively, and $A_{IB}$ represents the mixed coupling
of isospin and baryon number, and $g$ is the strong force charge.
 Let us assume that the constants $A_{B}$, $A_I$ and $A_{IB}$ are positive.
Taking into account the homogeneity of the universe we may disregard the
distinction between $i$ and $j$ and the formula becomes simplified somewhat,
\begin{eqnarray}
 V(r,N,Z)& = & \left( A_{B}(N+Z)^{2} + A_{I}(N-Z)^{2}
+ 2A_{IB}(N+Z)(N-Z)\right)g^{2}\frac{\exp{(-r/\lambda)}}{r}.\end{eqnarray}

\par If we assume that we are dealing with a conservative field, the
superweak force is given by minus the  derivative of the above
potential with respect to $r$, which is a function of time(along the
expansion).  Let us, at this point, deal with $r$ only.  We will discuss
time later on in connection with the expansion rate and the Hubble
constant.  Let us consider that the baryon number is conserved(and is the
same) in the two portions of matter subjected to the potential above, i.e.
$N+Z=B=constant$. Taking this into account the superweak force between the two
portions of the universe under consideration becomes
\begin{eqnarray}
 F(r,N(r),B)& = & -4\frac{dN(r)}{dr}\left(A_{I}(2N(r)-B)
+ A_{IB}B\right)g^{2}\frac{\exp{(-r/\lambda)}}{r} +
\nonumber \\
         &   & + \left(\frac{1}{\lambda} + \frac{1}{r}\right)V(r,N(r),B)
\end{eqnarray}

\noindent
where $B$ is the baryon number of any of the two portions. The number of
baryons of these two portions has to be extremely large, otherwise we would
already have clearly identified this force on Earth.
\par With the above expression for the superweak force we will be able to
explain the expansion of the universe itself and we may be able to show that
it may have a cyclic behavior. It also explains the evolution
of galaxies and the flat rotational curves of spiral galaxies.
\par We assume that at $t=0$, in the `beginning' of the universe,  $N$ had
a certain value, $N_o$(we will talk about its value later on), which decreased
via the weak interaction(free neutron decay) up to a time called
$t_p$.  At this time $N$ reached its minimum value,
being  only 13\% of all baryons, the remaining 87\% being protons.  The halt
in neutron decay happened due to the formation of atoms because the remaining
neutrons were no longer free. These neutrons became bound in the nuclei of
helium and deuterium.  At some time afterwards galaxies were formed.  This
paper discusses later on the possibility of having quasars as the precursors
of galaxies.  Due to gravitational attraction the clumping continued on small
scales and stars were formed.  From this point on nuclear fusion took place
in the core of stars of all galaxies(or quasars) and the number of protons
began to decrease slowly. This is explained as follows: As the universe ages
the stars become white dwarfs, neutron stars and black holes(not observed yet).
During the aging process the core density of a star increases and the high
electron Fermi energy drives electron capture onto nuclei and free protons.
This last process, called neutronization$^{(25)}$, happens via the weak
interaction. The most significant neutronization reactions are:
\newline
\begin{itemize}
\item {Electron capture by nuclei,}
\end{itemize}
\begin{eqnarray}
& & e
{-} + (Z,A)\;\;\stackrel{W}{\longrightarrow} \;\;\nu_{e} + (Z-1,A),
\end{eqnarray}
\begin{itemize}
\item {Electron capture by free protons,}
\end{itemize}
\begin{eqnarray}
& & e {-} + p \;\;\stackrel{W}{\longrightarrow} \;\;\nu_{e} + n,
\end{eqnarray}
\noindent
where $W$ means that both reactions proceed via  charged currents of the
electroweak interaction.
\par Of course, neutronization takes place in the
stars of all galaxies, and thus, the number of neutrons increases relative to
the number of protons as the universe ages. For example, a white dwarf
in the slow cooling stage(for $T{\leq}10^7$K) reaches a steady proton to
neutron density of about 1/8, and takes about $10^9$ years to
cool off completely, which is a time close to the present age of the universe.
By then, most stars have become white dwarfs(or neutron stars and possibly
black holes).  At a later time, because of the action of the superweak force,
during the contraction of the Universe, the galaxies will merge to each
other and with further contraction their stars will be disrupted.  This will
liberate neutrons and protons of stars. The neutrons of white dwarfs and
neutron stars will become free and will decay via the weak interaction and
the number of protons will increase again(with respect to the number of
neutrons). Therefore, we expect to have $\eta(r)=N(r)/B$ as a function of the
separation between the two bodies as shown in Fig.1. The separation when $\eta$
has a minimum is called $r_p$(p for protons); $r_n$ is the separation when
$\eta$ has a maximum(n for neutrons); when the separation is $R$ the
superweak force is zero. During the
contraction there must exist a separation when the contraction stops. This is
indicated by $r_o$. At this separation we will begin to count time. This is all
we need to have a cyclic Universe. In this way the superweak force will drive
the expansion and contraction of the universe and behaves overall as a
spring-like force(see Fig.2).  In Fig.2 $r_M$ and $r_m$ are the separations
when the force has a maximum and  a minimum, respectively. The other points
have been defined above. At present the separation $r$ must be larger than
$r_p$ and smaller than $R$. According to the Oort limit$^{(26)}$, roughly
half of the mass of our galaxy has already been processed by stars that have
completed their evolution. This means that there are many neutron stars and
white dwarfs which have not yet been detected in our galaxy. Since the same
must hold for all galaxies, the ratio of the number of neutrons to the number
of protons for the whole universe at the present epoch must be much larger
than 13/87. We can find some relations among the coupling constants from the
known number of neutrons and protons at some particular times.
\par In order to have a cyclic Universe the function $F(r)$ must have two
zeros, one corresponding to a stable position and the other one corresponding
to an unstable position. Let us analyse the stable position, first. Let us call
the time at this point, $T$ and the respective distance, $R$.
Let us expand the potential around $t=T(r=R)$ and
find the condition for a minimum(in the potential). Up to fourth order in
$r-R$ the potential is given by
\begin{eqnarray}\frac{V(r)}{(Bg) 2}&=&\frac{1}{R}\left(A_{B} +
2A_{IB}(2\eta_{T} - 1) + A_{I}(2\eta_{T} - 1)^{2}\right) +
\frac{1}{R}\left(4a_{T}A_{IB}\right.
\nonumber \\
& &\left. + 4a_{T}A_{I}(2\eta_{T} - 1)\right)(r-R)
 + \frac{1}{R}\left(4b_{T}A_{IB} +
4{a_{T}}^{2}A_{I} + 4b_{T}A_{I}(2\eta_{T} - 1)\right)(r-R)^{2}
\nonumber \\
& & + \frac{1}{R}\left(4c_{T}A_{IB} + 4c_{T}A_{I}(2\eta_{T} - 1)
+ 8a_{T}b_{T}A_{I}\right)(r-R)^{3}
\nonumber \\
& & + \frac{1}{R}\left(4d_{T}A_{IB}
+ 4{b_{T}}^{2}A_{I} + 4d_{T}A_{I}(2\eta_{T} - 1) +
8a_{T}c_{T}A_{I}\right)(r-R)^{4}
\end{eqnarray}

\noindent
where $a_T$, $b_T$, $c_T$ and $d_T$ are the first, second, third and
fourth derivatives of $\eta(r)$ with respect to $r$.
The linear term in $r-R$ should be zero so that we have a minimum at
$r=R$. This leads to the condition
\begin{eqnarray}
\eta_{T}&=& \frac{1}{2}\left(1 - \frac{A_{IB}}{A_I}\right).
\end{eqnarray}
\noindent
With this condition the potential becomes
\begin{eqnarray}
\frac{V(r)}{(Bg) 2}&=&\frac{1}{R}\left\{\left(A_{B} - A_{I}(2\eta_{T}
- 1)^{2}\right) +
{a_{T}}^{2}A_{I}(r-R)^{2}\right.
\nonumber \\
& & \left.+ 8a_{T}b_{T}A_{I}(r-R)^{3}
+ 4A_{I}(b_{T} {2} + a_{T}c_{T})(r-R)^{4}\right\}.
\end{eqnarray}
\noindent
Therefore, taking these conditions into account we obtain a pendulum-like
potential(and a restoring force at the bottom of the potential). Making the
first term equal to zero(just a reference level) we obtain that
$A_{B}A_{I} = A_{IB}$. Therefore, we may write $V(r)$ as
\begin{eqnarray} \frac{V(r)}{(Bg)^{2}} &=& 4A_{I}\frac{\left(\eta -
\eta_{T}\right)^{2}}{r}
\end{eqnarray}

\noindent
where $\eta_{T} = \frac{1}{2}(1 - \sqrt{\frac{A_B}{A_I}})$.
\par Thus, the expression for the force is given by
\begin{eqnarray} \frac{F(r)}{(Bg)^{2}} &=& - 8A_{I}\frac{\left(\eta -
\eta_{T}\right)}{r}\frac{d\eta}{dr} + 4A_{I}\frac{\left(\eta -
\eta_{T}\right)^{2}}{r^2}.\end{eqnarray}

\par We clearly see that $F=0$ when $\eta=\eta_{T}(t=T)$. In order to not
have three zeros for $F(r)$
\begin{eqnarray} \frac{d\eta}{dr} &{\neq}& \frac{\eta -
\eta_{T}}{2r}.\end{eqnarray}

\noindent
At $\eta=\eta_{p}$, $F$ is
\begin{eqnarray} \frac{F_{p}}{(Bg)
{2}} &=& 4A_{I}\frac{\left(\eta_{p} -
\eta_{T}\right)^{2}}{r^2}\end{eqnarray}

\noindent
which is positive, of course.
It is easy to show that the maximum in $F(r)$
happens at some point before $r=R$. When $F(r)$ is maximum
$\frac{dF(r)}{dr}=0$. Therefore, we obtain
\begin{eqnarray} 2(\eta - \eta_{T})\frac{d^{2}\eta}{dr^{2}} +
2\left(\frac{d\eta}{dr}\right)^{2} - 4\frac{(\eta -
\eta_{T})}{r}\frac{d\eta}{dr} + \frac{\left(\eta -
\eta_{T}\right)^{2}}{r^2} &=& 0 \end{eqnarray}

\noindent
which may be transformed into
\begin{eqnarray} 2(\eta - \eta_{T})\frac{d^{2}\eta}{dr^{2}} +
\left(\frac{d\eta}{dr}\right)^{2} + \left(\frac{d\eta}{dr} -
2\frac{(\eta - \eta_{T})}{r}\right)^{2} &=& 0. \end{eqnarray}

\noindent
The last two  terms are always positive. For $\eta<\eta_{T}$
$\frac{d^{2}\eta}{dr^{2}}$ has to be positive which means that the maximum
happens for $\eta<\eta_T$. The maximum may happen at $\eta=\eta_{p}$ if
\begin{eqnarray} \frac{d {2}\eta}{dr^{2}} &=& \frac{\eta_{T} -
\eta_{p}}{2r^{2}}.\end{eqnarray}

\par Of course, $R$ is the largest distance between the two bodies(galaxies,
or voids) and the Universe will have its maximum radius for  $\eta=\eta_{R}$.
\vskip .25in
VI. THE UNIVERSE OF ANTIPARTICLES
\vskip .15in
\par In order to have a contracting Universe the superweak force must become
negative. This is possible for times after the point $r=R$ only if we perform a
CP transformation in the relevant variables, $\vec{r}$ and $B^{2}{\eta}$. By
doing so we obtain a contracting Universe  of antiparticles.
\par Under a CP transformation $V(r)$ is left unchanged and $\vec{F}(r)$
goes into $-\vec{F}(r)$(because $\vec{\nabla}$ changes into $-\vec{\nabla}$).
In the contracting Universe let us name the force $\bar{F}(r)$.  Therefore,
$\bar{F}(r)$  given by
\begin{eqnarray} \bar{F}(r) &=& 4A_{I}(Bg)^{2}\left(2\frac{(\eta -
\eta_{T})}{r}\frac{d\eta}{dr} - \frac{(\eta -
\eta_{T})^{2}}{r^{2}} = - F(r)\right)\end{eqnarray}

\noindent
where we have used the fact that $\bar{\eta}=\bar{N}/\bar{B}=N/B=\eta$ in which
the bar indicates the antiparticle.
\par Because of the continuity of $F(r)$ at $r=R$ we must have
\begin{eqnarray} \frac{d\eta}{dr} _{\eta=\eta_{R}} &=& \frac{\eta_{R} -
\eta_{T}}{2R} = - F(r).\end{eqnarray}

\noindent
That is, we obtain $F(R)=0$.
\par For $\eta=\eta_n$ $\bar{F} _{\eta_n}= \bar{F}_{n}$ is given by
\begin{eqnarray} \bar{F} &=& - 4A_{I}(Bg)^{2}\frac{(\eta_{n} -
\eta_{T})^{2}}{{r_n}^2}\end{eqnarray}

\noindent
which is always negative. From Eqs. 16 and 17 we see that for $\eta=\eta_n$ the
second derivative of $\eta$ with respect to $r$ is negative since
$\eta_{n}>\eta_{T}$.
\par For $\eta=\eta_{R}$ the velocity, $v$, must be zero and for later times it
must become negative. Let us, now, analyse how the arrow of time is in the
contracting Universe. Under a CP transformation $\vec{v}$ changes sign which is
consistent with the fact of having a contraction instead of an expansion. This
means that we can not have a T transformation, that is, there is only one arrow
for time(otherwise, $\vec{v}$ would not change sign). Or, in other words,
time goes from zero(when the expansion begins) to a maximum value $T_{U}$(when
the contraction ends) and never has a negative rate(i.e., $dt$ is always
positive). $T_{U}$ represents a full cycle of the Universe.  Eqs. 6 and 7
are modified by  changing  particles into antiparticles.
\vskip .25in
VII. UNIFICATION OF THE SUPERWEAK FORCE WITH THE STRONG FORCE
\vskip .15in
\par Around  $t=0$, that is, for $r{\approx}r_{o}(\eta{\approx}\eta_{o}$ the
potential is given by
\begin{eqnarray} V(r) &=& (Bg)^{2}\frac{\left(\eta_{o} -
\eta_{T}\right)^{2}}{r}{\propto}\frac{B^{2}g^{2}}{r}
\end{eqnarray}
\noindent
which is an expression of the strong force potential. {\it{Therefore, the
strong force and the superweak force are unified at}} $t=0$.
\vskip .25in
\noindent
VIII. A POTENTIAL FOR THE UNIVERSE
\vskip .2in
\large
\hskip 4in{\it{ad infinitum}}
\normalsize
\vskip .15in
\par We represent in Fig.3  the potential of the superweak force according to
our calculations and considerations.  According to this figure  the universe
spends most of its time at the bottom of the potential, where it is more
stable. This time corresponds to the era of galaxies which are the units that
fill the Universe the longest.
\par The scale is not linear and that is why the curve has a smooth
descent and a  smooth rise. Actually, the potential is almost a square well
because the time $r_{p}$ is extremely small compared to $R$. Notice that $r$
diminishes after reaching the value $R$.
\par What is time in such a Universe? Time has a meaning only in each
cycle, for if the number of cycles is infinite the Universe is eternal. If we
consider that the number of cycles is finite, than we include, in fact, an
external agent to the Universe. This is a possibility, but in this case, we
would have other Universes.
\vskip .25in
\noindent
IX. TIME VARIATION OF THE HUBBLE CONSTANT
\vskip .15in
\par Let us now consider the variation of the Hubble constant with time(during
the expansion).
According to our arguments, it must be decreasing with time since $t=t_p$(which
corresponds to $r=r_p$), and will continue decreasing up to $t=t_n$(which
corresponds to $r=r_n$. In other words, the expansion must be slowing down at
present.
\par As is well known, it is convenient to represent the universal expansion
rate as
\begin{eqnarray} \displaystyle \frac{1}{R(t)}\frac{dR(t)}{dt} & = & f(t)=H(t)
\end{eqnarray}

\noindent
where $H(t)$ is  Hubble's constant. For an expanding universe
$H(t)$ is positive.  The velocity between two bodies(galaxies, for example)
separated by a distance $r(t)$ is given by
\begin{eqnarray} v(t) & = & H(t)r(t). \end{eqnarray}

\par The relative velocity between any pair of galaxies(or the galactic
superstructures) is a rather small velocity. Therefore we may use
classical Newtonian mechanics. Having this in mind let us  consider
that the mass of a body is, to a very good approximation, given by
$m{\approx}Bm_p$, and let us make ${\eta}=N/B$. The force between two
bodies($m_{1}=m_{2}$, $B_{1}=B_{2}$ and $N_{1}=N_{2}$)  will be given by
\begin{eqnarray} m_{p}\frac{d^{2}r}{dt^2} & = & - \frac{1}{r}\frac{dL}{dr}
+ \frac{L}{r^2}\end{eqnarray}

\noindent
where
\begin{eqnarray} L(\eta) & = & 4A_{I}(\eta - \eta_{T})^{2}(Bg)^{2}.
\end{eqnarray}

\noindent
We may make $r(t)=R(t)r_o$, where $r_o$ is the initial separation between
the two bodies and writing $r(t)$ in terms or $R(t)$ and $H(t)$
one obtains
\begin{eqnarray} \dot{H} & = & - \frac{\dot{L}(\eta)}
{m_{p}{r_o}^{3}HR^3} +
\frac{L(\eta)}{m_{p}{r_o}^{3}R^3} - H^2. \end{eqnarray}

\noindent
Let us recall that  $\dot{L}(\eta)$ is given by
$\dot{L}(\eta)=8A_{I}(Bg)^{2}\dot{\eta}(\eta - \eta{T})$. In the range between
$t=t_p$ and $t=t_T$,  $\dot{\eta}>0$ and $\eta<\eta_T$, and therefore,
$\dot(L)$ is negative. Let us consider $t$ as beeing the present epoch of the
Universe. If the expansion is slowing down we must have $\dot{H}<0, L>0$
in this range.  Solving the cubic  equation in $\dot{H}$ for $\dot{H}<0$, we
obtain
\begin{eqnarray} H(t) & > & \left(\frac{ \dot{L} }{2m_{p}{r_o}^{3}R^3}
+ \sqrt{-\frac{L^{3}}{27{m_p}^{3}{r_o}^{9}R^9} +
\frac{(\dot{L})^{2}}{4{m_p}^{2}{r_o}^{6}R^6}}\right)^{1/3}
\nonumber \\
& & - \left(\frac{ \dot{L} }{2m_{p}{r_o}^{3}R^3} -
\sqrt{-\frac{L^{3}}{27{m_p}^{3}{r_o}^{9}R^9} +
\frac{(\dot{L})^{2}}{4{m_p}^{2}{r_o}^{6}R^6}}\right)^{1/3}
\end{eqnarray}

\noindent
which means that $H$ is positive.
\vskip .15in
X. A NEW POSSIBLE BEGINNING FOR THE UNIVERSE
\vskip .25in
\par We can calculate how $H(t)$ behaves with time around $t=0$ if we make some
assumptions on the relative proportions of neutrons to protons prevailing
around $t=0$.  The nuclear reactions which must be considered in determining
the proton-neutron ratio are the following:
\begin{eqnarray}
& & n \rightleftharpoons p + e {-} + \bar{\nu}_{e}, \nonumber
\end{eqnarray}
\begin{eqnarray}
& & n + \nu_{e} \rightleftharpoons p + e {-}, \nonumber
\end{eqnarray}
\begin{eqnarray}
& & n + e {+} \rightleftharpoons p + \bar{\nu}_{e}. \nonumber
\end{eqnarray}

\noindent
By considering that the temperature is {\it{not}} very high  so that
$m_{e}c^{2}{\approx}kT$, Alpher et al.$^{(27)}$ have shown(in another context)
that, among the reactions above, free neutron decay is the dominant reaction.
If the neutrons are free than $\eta(t)=\eta_{o}\exp{(-t/\tau)}$ and
\begin{eqnarray} \frac{d\eta}{dt} _{t=0} &=& \eta_{o}.\end{eqnarray}
\noindent

In order to have a cyclic Universe $v$ must be zero at two different times
which are $t=0$ and $t=t_R$. But, because $\frac{d\eta}{dt} =
\frac{d\eta}{dr}v$, the neutrons can not be completely free at $t=0$, i.e..
$\frac{d\eta}{dt}$ can not be exactly equal to a constant. For times around
$t=0$ \hskip .2in $\eta(t)$  must be
\begin{eqnarray} \eta(t) &=& \eta_{o}\exp{(-g(t))}\end{eqnarray}

\noindent
such that, for times close to zero, $g(t){\propto}t^{x}$ where $x>1$.
This will assure that $H(t=0)=0$. Therefore, there must exist a process that
will speed up the decay of neutrons.
\par The above condition concerning the temperature at $t=0$ is included in
the following considerations about the temperature of the universe at some
particular times:
\begin{itemize}
\item{Each cycle of the universe begins($t=0$) with a volume of neutrons,
protons and electrons at a temperature of about $1$Mev, which
is necessary for the primordial formation of the light elements. On this
we must reanalyse the papers of Gamow$^{(28)}$ and of Alpher and
Herman$^{(29)}$ taking into account this new interaction.}
\item{At $t=t_p$ the temperature is about $0.1$Mev(necessary because of the
deuteron binding energy). Considering free neutron
decay this corresponds to $t_{p}=25min$.} As we considered above the neutrons
must decay at a faster rate and therefore $t_{p}<25min$.
\item{The atom is formed when the temperature has dropped to about $1ev$. This
is consistent because for an adiabatic expansion $T{\sim}l^{-1}$, and since
the hydrogen atom is $10^5$ larger than a nucleon, we expect the
temperature to be about $0.1{\times}10^{-5}$Mev$=1$ev.}
\end{itemize}
\par This possible history of the universe is consistent with the observed
quantity of matter with respect to antimatter.  What has been observed is that
antimatter does not exist by itself and only comes out from nuclear
reactions(involving matter) which take place in stars or proto-stars.
\par This history of the Universe is also consistent with quark confinement.
As is well known all high energy experiments have shown that quarks are
`mysteriously' confined. If they are permanently confined, then they were
never free.  That is exactly what this paper assumes, i. e., the paper
assumes that neutrons and protons always existed.  In this fashion quark
confinement is linked to the existence of the superweak force and may be linked
to a conservation in the number of nucleons.
\par In the light of the above considerations we must reinvestigate the origin
of the cosmic black body radiation. Because of the recent observations on
this radiation$^{(4)}$ we need to reinvestigate its origin anyway. COBE'S  data
show that the radiation is not absolutely isotropic and exhibits dipole and
quadrupole moments. As has been considered by several
investigators$^{(30,31,32,33,34,35)}$, this radiation may have been produced by
a generation of `population III' stars which were formed and burned out quickly
before galaxy formation.

\vskip .25in
\noindent
XI. QUASARS AS THE PRECURSORS OF GALAXIES
\vskip .15in
\par
A typical galaxy has $10^{11}$ suns which correspond, roughly, to $10^{68}$
hydrogen atoms. When we consider these atoms close to each other they occupy
a volume with a radius of the order of $5{\times}10^{14}$cm${\approx}0.1$ light
day. This figure is a lower bound for the radius of a quasar. We will see below
that their radii have to be larger than 10 light years. The radius of the bulge
of our galaxy is of the order of $10^{4}$ light years. Also, the velocity $v$
at the equator of a quasar has to be smaller than the speed of light.
Therefore, using angular momentum conservation we obtain that
$R_{Q}> R_{B}v_{B}/c$, where $R_{Q}$ is the radius of the quasar, $R_{B}$ is
the radius of the bulge of our galaxy and $v_{B}$ is the velocity of a star
at the equator of our galaxy's bulge. One finds that $R_{Q}> 10$ light years.
This is an increase of $10^4$ over the radius previously calculated. The
corresponding upper bound for the temperature is $10^{-4}$eV. The radius of
$0.1$ light day gives a density of the order of the density of water. The
upper bound, $10$ light years, provides as a lower bound for the density the
figure $10^{-12}$g/cm$^3$. This means that quasars were formed after the
formation of hydrogen atoms. Therefore, they were formed at a temperature
between $0.1$mev and $1ev$. At this temperature the kinetic energy of the gas
particles at the surface of the gas cloud is larger than the gravitational
potential of the gas particles.  This is not a problem because each gas
cloud(quasar) will expand anyway due to action of the superweak force. In
this picture quasars are formed due to the action of the superweak force. Since
this force has an infinite range the repulsion must also exist on larger
scales.
\par We may find that quasars have radii around 10 light years in another way.
It is well known that they have an angular diameter of ${\theta}=10^{-3}$
arcsec. The most distant quasars are at distances of the order of $cT$, where
$T$ is the present age of the Universe. Therefore, their diameters are
$d = cT{\theta}{\approx}10$ light years. It is known that some quasars vary
their brightness from night to night. This flickering may come just from their
brighter centers and does not invalidate the above figure of 10 light years. Of
course, quasars with lower masses must also have smaller radii.
\par As is well known we find quasars at redshifts beyond those to which we can
see galaxies and there is no quasar at small redshifts. It is shown
below how they may have evolved into galaxies. There is also a sharp drop in
their numbers at a redshift around $z=2.2^{(36)}$.  This redshift is just a
mark showing when they were formed.
\vskip .25in
\noindent
XII. THE EVOLUTION OF GALAXIES
\vskip .1in
\parbox{3in}{a capite ad calcem}
\vskip .15in
\par A new born quasar, as discussed above, must have most of
its mass as hydrogen, the rest being the primordial helium.  Therefore, its
hydrogen atoms experience a repulsion among themselves because of the
superweak force. The repulsive force between two hydrogen atoms separated by a
distance $d$ is given by
\begin{eqnarray} F_{pp} & = & \frac{\left(\sqrt{A_{B}} - \sqrt{A_{I}}
\right)^{2}}{d^{2}}g^{2}\end{eqnarray}.

\noindent
This repulsion among all hydrogen atoms makes the quasar increase in size and
go through the intermediate stages which may include radio galaxies and BL
Lacertae objects.
The big clumps of hydrogen and helium gasses form the globular
clusters(and the small clumps in each cluster form its stars).  There must
also exist repulsion between each globular cluster and the galactic nucleus(and
among the globular clusters). From this point on the evolution may follow three
different branches:
\vskip .2in
\noindent
a) Elliptical galaxy.
\vskip .15in
\par A quasar may become an elliptical galaxy by expanding slowly as a whole.
Because of rotation we may have several types of ellipticals according to
their oblateness. As is well known ellipticals do not exhibit much rotation(as
compared to spiral galaxies). This is explained as follows: As the quasar
expands its angular velocity decreases because of angular momentum
conservation. For example, the angular velocity of an EO must be given
by(disregarding mass loss)
\begin{eqnarray} {\omega}_{EO} &=& {\omega}_{Q}{\frac{R_{EO}}{R_{Q}}}^{2}
\end{eqnarray}

\noindent
where $\omega_Q$ is the angular velocity of the quasar which gave origin to the
galaxy; $R_{EO}$ and $R_{Q}$ are the radii of the elliptical galaxy and the
quasar respectively. Because $R_{EO}{\gg}R_{Q}$,
${\omega}_{EO}{\ll}{\omega}_{Q}$. This is consistent with the slow rotation
of ellipticals.  There is also the following consistency to be considered. Most
galaxies in the Universe are ellipticals(about 60\%) and as was shown above
this means that most quasars expand slowly. Therefore, most quasars must not
show rapid variability and must also be radio quite.  This is exactly what
has been reported$^{(37)}$.
\par Another evidence to be taken into account is the reported nebulosity
around some quasars.  Boroson and Person$^{(38)}$ have studied this nebulosity
spectrocopically.  The emission lines they found are of the same type as the
emissions from a plasma.
\vskip .2in
\noindent
b) Spiral galaxy.
\vskip .15in
\par There are two possibilities in this case: normal spiral and barred spiral.
This happens when, at some point in its expansion towards becoming a galaxy,
a quasar expands rapidly by pouring matter outwards from its center. This
pouring will give origin to two jets which will wind up around the central
bulge because of rotation and will create the two spiral arms. A possible
mechanism is the following: Due to rotation we expect to have some bulging in
the spherical shape, and because of angular momentum conservation  the
outpouring of matter may only happen in a plane perpendicular to the axis of
rotation.  Because of rotation the core of the quasar becomes also an
ellipsoid.  This core(which has a higher concentration of baryons) may be
broken in two parts, going to a state of lower potential energy(of the
superweak interaction). These two parts repel each other and form two
centers(lobes) in the equator of the quasar(or young galaxy). These two lobes
are seen in many radio galaxies. Because of the outpouring of matter from each
center, there must exist all kinds of radiation covering the whole
electromagnetic spectrum especially in the form of synchroton radiation caused
by collisions among atoms.  Because of these collisions we expect to have
electrons stripped from  hydrogen and helium atoms.  These electrons will
create the observed synchroton radiation which is associated with jets in
very active galaxies.  These collisions provide also the enormous energy
output observed in quasars.
\par  As the galaxy ages its bulge diminishes and leaves the globular
clusters alone without the embedding gas since the gas has either escaped
or has been transformed into globular clusters.  Also, the activity at the
galactic center diminishes as the age increases due to the increase in the
number of neutrons(which decreases the superweak force) because of nuclear
fusion and also because of the decrease in material at the nucleus.
\par  The barred spirals are galaxies that have expelled matter more
vigorously. That is exactly why their arms are not tightly wound.
As the galaxy ages the arms will curl up more and more and the bar will
disappear because of the ejection of material outwards.  It is worth
noting that the more spirals(including barred ones) are wound up, the smaller
are their nuclei and, conversely, the larger are their bulges, the less they
are wound up.  This happens because of the shedding of matter outwards from
their  nuclei throughout the galaxy's life due to the action of the superweak
force. The bar can be explained in terms of a more vigorous shedding
of matter outwards as compared to the shedding that takes place in normal
spirals. Therefore, as a spiral evolves its nucleus diminishes and the two
arms become more and more tightly wound up. Let us now show why the core
preferentially breaks up in two parts.  Let us compare the potential energies
when the core is broken in two and three parts.  For simplification, let us
consider that the core is broken in equal parts. In the case of two parts we
have
\begin{eqnarray} V_{2} & = & \frac{1}{4}\frac{q^{2}}{d_{2}}\end{eqnarray}

\noindent
and in the case of three parts,
\begin{eqnarray} V_{3} & = & \frac{1}{3}\frac{q^{2}}{d_{3}}\end{eqnarray}

\noindent
where $q=\left(\sqrt{A_B}B + \sqrt{A_I}(N - Z)\right)g^{2}$, in which $B$, $N$
and $Z$ are the numbers of baryons, neutrons and protons of the core before
it is broken.  Clearly, $V_{2}<V_{3}$, and therefore the breaking in two
parts is preferred for it corresponds a state of lower energy.
\par Since a quasar must consist mainly of hydrogen gas, let us disregard the
number of neutrons.  In this case the potential between the two parts(separated
by d) is given by
\begin{eqnarray} V = g {2}\left(\sqrt{A_B} - \sqrt{A_I}\right)^{2}
\frac{B_{1}B_{2}}{d}\end{eqnarray}
\noindent

in which $B_{1}$ and $B_{2}$ are the number of baryons in the two parts and
$B_{1} + B_{2}=B(constant)$. This is consistent with the observation of the two
lobes of radio galaxies. For example, a contour map of radio emissions shows
that most of the radiation comes from two blobs located on either side of the
central galaxy Cygnus A(also called 3C405). When observed in the visible
spectrum the central galaxy also shows two blobs. In the framework of gravity
it is impossible to explain the situation of this galaxy and its four blobs.
They must have been formed due to the action of the superweak force. At some
time in the past there was just a single agglomeration of gas which separated
in two main parts leaving a small part in the middle. The radio emission must
be caused by the outpouring of matter which takes place in each blob.
\vskip .2in
\noindent
c) Seyfert galaxy.
\vskip .15in
\par In this case the expansion is more violent. It is well known that these
galaxies are powerful sources of infrared radiation.  Several Seyfert
galaxies are also strong radio emitters, and also, like quasars, they
present variability in their energy output.  For example,
over a period of a few months, the nucleus of the Seyfert galaxy M77(or
NGC1068) switches on and off a power output equivalent to the total luminosity
of our galaxy. It is also worth noting that the nuclei of Seyfert galaxies
are very bright and have a general starlike appearance.
\par Researchers have found that Seyfert galaxies exhibit explosive
phenomena$^{(39)}$. For example, M77 and NGC4151 expel huge amounts of gas from
their nuclei.  The spectra of both galaxies show strong emission lines, just as
quasars'.
\par In summary, the proposed evolution for galaxies including all kinds of
galaxies follows the two branches: \newline
\vskip .2in
I) Quasar(without jets) $\rightarrow$ (BL Lacertae or radio galaxy)
$ \rightarrow \left\{ \begin{array}{l}
                          \mbox{Seyfert Galaxies} \\
                          \mbox{Elliptical Galaxies} \\
                          \mbox{Spiral Galaxies}
                          \end{array}
               \right. $
\vskip .2in
II) Quasar(with jets) $ \rightarrow$ radio galaxy
$ \rightarrow \left\{ \begin{array}{l}
                       \mbox{Seyfert Galaxies}\\
                       \mbox{Spiral Galaxies}
                       \end{array}
                                      \right. $
\vskip .2in

\par The BL Lacertae objects stage may or may not be present.  Radio galaxies
represent an intermediate stage. Irregular galaxies have not been included
but they only represent 10\% of all galaxies and they are not well understood.
Besides, it is suspected that many irregulars are the result of collisions
among galaxies.
\par Of course, there must exist evolution in each final branch.  As
ellipticals age they become more and more oblate as a result of rotation.
Spiral galaxies evolve by means of the shedding of matter from their centers
towards the disk.  Seyfert galaxies may display a complex evolution since not
all of them exhibit spiral arms. It is clear that as these galaxies age their
nuclei will be less and less bright.
\par Let us, now, make a general analysis including all kinds of galaxies.
Considering the evolution above proposed we do not expect to have very small
spiral galaxies because spirals must come from strong expulsion of matter from
quasars nuclei, and this must happen only when the number of baryons is
sufficiently large.  This is the case, indeed, because  dwarf galaxies are
either irregular or elliptical galaxies.  Ellipticals have masses that range
between $10^{5}$ and $10^{13}$ solar masses while spirals masses are
comprised between $10^{9}$ and $10^{11}$ solar masses.  Also, we expect that
spirals have faster rotations than ellipticals and, indeed, they do.  This
is just because the nuclei of spirals are smaller than ellipticals galaxies(for
the same mass, of course). Therefore, according to Eq.30 above, spirals should
have faster rotational velocities.  It is also expected that, since spirals
shed gas in their disk throughout their lifetimes, their disks must have young
stars.  This is an established fact.  Our galaxy's disk, for example, has very
hot, young(O-,B-, and A-type) stars, type-I Cepheids, supergiants, open
clusters, and interstellar gas and dust.  Each of these types represent young
stars or the material from which they are formed. Conversely, the globular
clusters and the nucleus contain older stars, such as RR Lyrae, type-II
Cepheids, and long-period variables.  This, of course, is a general
characteristic of all spirals.  For example, Young O- and B-type stars are the
stars which outline the beautiful spiral arms of the Whirlpool galaxy. Because
of the lack of gas(i.e., because of the lack of a disk) ellipticals also have
primarily very old stars.
\par It is well known that BL Lacertae objects are powerful sources of radio
waves and infrared radiation.  They share with quasars the fact of exhibiting
a starlike appearance and of showing short-term brightness fluctuations.  As
some quasars do, they also have a nebulosity around the brighter nucleus.
Researchers$^{(39)}$ have managed to obtain the spectrum of their nebulosity.
{\it{The spectrum of the nebulosity is strikingly similar to the spectrum of
an elliptical galaxy}}(M32's spectrum, in this case). Because of the
short-term variability these objects are very compact objects. In terms of
the evolution above described they are simply an evolutionary stage of a
quasar towards becoming a normal galaxy.
\par Radio galaxies share with BL Lacertae object many of the properties of
quasars. As Heckman et al.$^{(40)}$ have shown, in the middle and far
infrared(MFIR) quasars are more powerful sources of MFIR radiation than radio
galaxies. Also, there have been investigations showing that the emission from
the narrow-line region(NLR) in radio-loud quasars is stronger than in radio
galaxies of the same radio power$^{(41,42,43)}$. Goodrich and
Cohen$^{(44)}$ have studied the polarization in the broad-line radio galaxy
3C 109. After the intervening dust is taken into account the absolute
V-magnitude of this galaxy becomes $-26.6$ or brighter, which puts it in the
quasar luminosity range. The investigators suggest that ``many radio galaxies
may be quasars with their jets pointed away from our direct line of sight''.
It has also been established that radio galaxies are found at intermediate
or high redshifts and that they are clearly related to galactic evolution
because as the redshift increases cluster galaxies become bluer on average,
and contain more young stars in their nuclei. This is also valid for radio
galaxies: the higher the redshift, the higher their activity. All these data
show that a radio galaxy is just an evolutionary stage of a galaxy towards
becoming a normal galaxy, i.e., it is just a stage of the slow transformation
by means of an overall expansion of a quasar into a normal galaxy.
\par In the light of the above considerations the nuclei of old spirals must
exhibit a moderate activity.  This is actually the case.  The activity must
be inversely proportional to the galaxy's age, i.e., it must be a function of
luminosity.  The bluer they are, the more active their nuclei must be.  As
discussed above there must also exist a relation between this activity and
the size of the nucleus(as compared to the disk) in spiral galaxies.  Our
galaxy has a mild activity at its center.  Most of the activity is
concentrated in a region called Sagittarius A, which includes the galactic
center.  It emits synchroton and infrared radiations.  Despite its large energy
output Sagittarius A is quite small, being only about 40 light years in
diameter.  Besides Sagittarius A our galaxy exposes other evidences showing
that in the past it was a much more compact object: 1) Close to the center,
{\it{on opposite sides of it}}, there are two enormous expanding arms of
hydrogen going away from the center at speeds of 53km/s and 153km/s; b) Even
closer to the center there is the ring called Sagittarius B2 which is expanding
at a speed of 110km/s$^{(45)}$. It is worth noting that the speeds are very
low(as compared to the velocities of relativistic electrons from possible black
holes). This expansion is just a manifestation of the action of the superweak
force. This is not restricted to our galaxy. Recent high-resolution
molecular-line observations of external galaxies have revealed that galactic
nuclei are often associated with similar expanding rings$^{(46)}$. When the
galactic center is seen at radio wavelengths(3.75cm) it shows a flattened shape
along the galactic equator.  This fact also lends support to the above
considerations.
\par The nearby galaxy M31 provides some valuable data towards the evolutionary
scheme above discussed.  The nucleus of this galaxy is a strong emitter of
infrared radiation, even though there is no known star formation in this
part of the galaxy.  This radiation may be explained in terms of the expansion
of gas from its center.
\par A very important result about galaxies is the very tight relationship
between the far-infrared and non-thermal radio emission that extends over
nearly three orders of magnitude$^{(47)}$.  Therefore, these two processes may
be tied together.  Our interpretation is that they are caused by the expulsion
of matter from the center. As discussed above this expulsion gives off lots of
synchroton and infrared radiation.
\par We may prove or disprove the above evolutionary hypothesis by testing
some  of the predictions of the evolution above displayed:
\begin{enumerate}
   \item The number of spirals must decrease with increasing redshift;
   \item The sizes of spirals disks must decrease when we go to higher
redshifts;
   \item The activity in the nuclei of spirals must be directly connected with
the sizes of their nuclei.  The larger their nuclei are, the higher must be the
activity in the nuclei(in terms of radio emission and infrared radiation);
   \item Small galaxies must hardly exhibit any activity in their nuclei;
 \end{enumerate}
\par A very important support to the above evolution is provided by the number-
luminosity relation $N(>l)$. When expanded in terms of the apparent luminosity,
$l$, the first term(Euclidean term) is  given by$^{(36)}$
\begin{eqnarray} N(>l) &=& \frac{4{\pi}n(t_{o}}{3}
{\left(\frac{L}{4{\pi}l}\right)}^{1.5}\end{eqnarray}

\noindent
where $n(t_{o})$ is the present density of galaxies and $L$ is the absolute
magnitude. The correction term is always negative, so that the number of
faint objects($l$ small) should always be less than the number that the
$l^{1.5}$ predicts. This conclusion is strongly contradicted by observations on
radio sources: many surveys of radio sources agree that there are more faint
sources than the $l^{1.5}$ law predicts.  The fitting of the experimental data
provides a law of the form$^{(36)}$
\begin{eqnarray} N(>l) {\approx} \frac{constan}{l^{1.8}}\end{eqnarray}

Since the formula breaks down for small $l$(i.e., faint distant sources), we
must conclude that in the past radio sources were brighter and/or more
numerous that they are today. This, of course, lends support to the above
evolutionary scheme.
\par Because the superweak force goes with the square of the number of
baryons we expect that quasars with small masses(i.e., small number of baryons)
must evolve less rapidly and more quietly.  Also, small ellipticals must be
very quiet galaxies.  This is a well known fact.
\par At this point it is worth mentioning that there is a very important
drawback against the traditional view of explaining the formation of arms in
spirals by the bulging effect of rotation. If this were the case we would find
a higher proportion of pulsars off the galactic equator of our galaxy. But the
real distribution reveals that these sources are mostly concentrated in the
galactic equator.  The traditional view does not explain either why all spirals
have large amounts of gas in their disks. Besides, within the traditional
framework quasars are just exotic objects. Evolution is clearly out of question
without a repulsive force. \vskip .25in
\noindent
XIII. THE ARMS OF SPIRAL GALAXIES
\vskip .15in
\par The rotational curve of spiral galaxies is one of the biggest puzzles of
nature. It is possible to give a reasonable explanation for this puzzle
in terms of the action of the superweak force.  In the process we will also
explain the formation of the spiral structure of the arms.
\par First, let us consider the central nucleus(or bulge).  The bulge was
formed by means of an overall repulsion mainly  among all hydrogen atoms.  This
repulsion took place a long time ago as the globular clusters show us.  For
simplification let us consider a uniform density for the bulge.  Because mass
varies as $r^3$ and the gravitational force varies as $r^{-2}$ we expect the
tangential velocity to be proportional to $r$.  This provides a cancellation
of the gravitational force with the centrifugal force.  The superweak force
just contributed to the small radial velocities at the time of formation of
the bulge.
\par Now, let us consider the tangential velocities of stars in the disk.
As was shown above the disk was formed by the shedding of matter from the
center of the galaxy where a denser core existed.  Let us consider that the
bulge has a radius $R_{B}$.  Let us consider that, because of the action of
the superweak force, a certain mass of gas $m$ is expelled from the center.
While the mass is inside the bulge its tangential velocity will increase
with $r$ and, therefore, will reach its maximum value, $v_{B}$, at $R_{B}$.
Because of its radial velocity, the mass $m$ will continue to distance itself
from the bulge.  But its tangential velocity is kept fixed because of the
action of repulsion and because of the tranfer of angular momentum from the
bulge to the mass.  This may be shown in the following way: As the mass goes
away from the center it increases its angular momentum. At a distance $r$ the
angular momentum is given by
\begin{eqnarray} L &=& mrv_{t}\end{eqnarray}

\noindent
where $v_t$ is the tangential velocity. Because $L$(of the mass $m$) increases
with time(and with $r$) we have
\begin{eqnarray} \frac{dv_{t}}{v_{t}} &>& -\frac{dr}{r}.\end{eqnarray}
\noindent

Integrating, we obtain
\begin{eqnarray} ln{\frac{v_{t}}{v_{to}}} &>& ln{\frac{r_{o}}{r}}\end{eqnarray}
\noindent

where $r_o$ is the position of the mass at a time $t_o$ and $r$ is its position
at a later time. Both positions are measured from the center. Because the
logarithm is an increasing function of the argument, we must have
\begin{eqnarray} \frac{v_t}{v_{to}} &>& \frac{r_o}{r}.\end{eqnarray}
\noindent

We clearly see that $v_{t} = v_{to}$ is a solution of the above inequality
because $r$ is always larger than $r_o$.
\par Now, let us examine what happens to the mass in terms of energy. At a
position $r$ from the center of the galaxy its energy is given by
\begin{eqnarray} E &=& \frac{L^{2}}{2mr^2} + \frac{1}{2}m{v_r}^{2} -
\frac{GMm}{r} + U(r)\end{eqnarray}
\noindent

where $L$ is the star's angular momentum, $v_r$ is the radial velocity of the
mass $m$, $M$ is the mass of the galaxy, $G$ is the gravitaional constant
and $U(r)$ is the repulsive potential of the superweak force. The first time
derivative of the energy(of the mass m) is
\begin{eqnarray} \frac{dE}{dt} &=& mv_{t}\frac{dv_{t}}{dt} +
mv_{r}\frac{dv_{r}}{dt} + \frac{GMm}{r^2}\frac{dr}{dt} +
\frac{dU}{dt}.\end{eqnarray}
\noindent

Assuming a conservative field and disregarding the action of any other forces
on the mass $m$, we must have $\frac{dE}{dt}=0$. Because the superweak force
is a radial force $\frac{dv_t}{dt}$ must be zero. Therefore, $v_t$ is a
constant and is, therefore, independent of the distance to the center of the
galaxy. The superweak potential $U(r)$ is given by
\begin{eqnarray} U(r) &=& \frac{\left(\sqrt{A_B} + \sqrt{A_I}\left(2\eta -
1\right)\right)^{2}Bbg^{2}}{r}\end{eqnarray}
\noindent

where we have considered that the bulge and the mass m have the same proportion
of neutrons to baryons, $\eta$. Let us call $Q=\left(\sqrt{A_B} +
\sqrt{A_I}(2\eta - 1)\right)$ to make the calculation simpler. Taking this into
account we have
\begin{eqnarray} \frac{dU}{dt} &=& \frac{2BbQ}{r}\frac{dQ}{dt} -
\frac{BbQ^{2}}{r^2}\frac{dr}{dt}.\end{eqnarray}
\noindent

Substituting this expression into Eq.(57), we obtain that
\begin{eqnarray} v_{r} &=&
\frac{\frac{2BbQ}{r}\frac{dQ}{dt}}{frac{BbQ^{2}}{r^2} + \frac{GMm}{r^2} -
m\frac{dv_{r}}{dt}}.\end{eqnarray}
\noindent

Let us analyse in detail the above expression for $v_r$. As the mass $m$ goes
further from the center the superweak force diminishes. Therefore,
$\frac{dv_t}{dt}$ must be negative. But, since the middle term of the
denominator is dominant(because the superweak force is weaker than the
gravitational force), the denominator  is always negative. The numerator
is given by
\begin{eqnarray} \frac{2BbQ}{r}\frac{dQ}{dt} &=& 2Bb\left(\sqrt{A_B} +
\sqrt{A_I}(2\eta - 1)\right)2\frac{d\eta}{dt}\sqrt{A_I}\end{eqnarray}
\noindent

which, with the use of Eq.(13), may be transformed into
\begin{eqnarray} \frac{2BbQ}{r}\frac{dQ}{dt} &=& 8BbA_{I}g^{2}(\eta - \eta_{T})
\frac{d{\eta}}{dt}. \end{eqnarray}
\noindent

As we saw before, for $T_{t}>t>t_{p}$, $\eta<\eta_{T}$ and
$\frac{d{\eta}}{dt}>0$. Therefore, $v_r$ is always positive, i.e., {\it{the
mass m moves outward and, as time goes by, it gets further from the center}}.
In this way the mass $m$ gains angular momentum. Because of conservation of
angular momentum the galactic nucleus must
decrease its angular momentum by the same amount.  If we consider
that the angular velocity of the nucleus does not diminish(which is more
plausible than otherwise), then its mass must diminish, i.e., the nucleus sheds
 more matter outwards.  Since $v_t$ remains the same the angular velocity must
decrease as the mass goes away from the center. This generates the differential
rotation observed in all spiral galaxies. The formation of the spiral structure
is, therefore, directly connected with the evolution of the galaxy. Fig.6.
illustrates the formation of the differential rotation.
\par We may easily show that the beautiful spiral arms are described by a
logarithmic spiral(in an inertial frame). For this let us consider
Fig.7.  In this figure the mass
{\it{m}} has moved away from the bulge following the curve C. The point P in
the bulge where the mass {\it{m}} passed has moved to the position Q. Also, we
have seen that the tangential velocity of the mass {\it{m}} is a constant.
Therefore,
\begin{eqnarray} r{\omega} &=& r\frac{d{\theta}}{dt} = R\frac{d{\phi}}{dt} =
R{\Omega} = v_{t} = constant \end{eqnarray}

\noindent
where $R$ is the radius of the galactic bulge and $v_t$ is the tangential
velocity of the mass $m$. We have that
\begin{eqnarray} d{\theta} &=& {\omega}dt =  \frac{R{\Omega}}{r}dt =
\frac{R{\Omega}}{rv_{r}}dr\end{eqnarray}

\noindent
where we have used the fact that $v_{r} = \frac{dr}{dt}$. Considering that
$v_r$ varies slowly with $r$(or $t$) we may integrate $d{\theta}$ and obtain
\begin{eqnarray} r &{\approx}& Re^{\frac{v_{r}}{v_{t}}\theta}\end{eqnarray}.

\noindent
{\it{This is the equation of the logarithmic spiral}}. We imediately obtain
that
\begin{eqnarray} \omega &{\approx}& {\Omega}e^{-\frac{v_r}{v_t}\theta}.
\end{eqnarray}

\noindent
We may also calculate $\phi$. It is given by
\begin{eqnarray} \phi &{\approx}& k\left(e^{\frac{\theta}{k}} - 1\right)
\end{eqnarray}

\noindent
where $k$ is given by $v_{t}/v_{r}$.
\par The ratio $k=v_{t}/v_{r}$  distinguishes between the two
types of spiral galaxies. If $k{\ll}1$, then $\omega$ diminishes rapidly with
$\theta$. This corresponds to spirals with
bars. Conversely, if $k{\gg}1$, then $\omega$ diminishes slowly and only
reaches a very low value for large $\theta$. This is consistent with the
experimental data on spiral galaxies. The middle ground $k{\approx}1$ must
correspond to intermediate cases.
\par From the point of view of a frame fixed in the galactic bulge and rotating
with it the mass $m$ describes an arch according to Fig.8. The angle $\psi$ is
related to $\theta$ and $\phi$ by $\psi = \phi - \theta$. Therefore, $\psi$
is given by
\begin{eqnarray} \psi &=& k\left(e^{\frac{\theta}{k}} - 1\right) - \theta.
\end{eqnarray}

\noindent
For small $\theta$ one has $\psi {\approx} {\theta}^{2}/{2k}$ and
\begin{eqnarray} r {\approx} Re^{\sqrt{\frac{2}{k}}{\psi}^{.5}}\end{eqnarray}

\noindent
and for large $\theta$ we have $\psi {\approx} ke^{\frac{\theta}{k}}$ and
\begin{eqnarray} r {\approx} R\frac{\psi}{k}\end{eqnarray}.

\noindent
Therefore, in this rotating frame the mass m also describes a spiral curve as
it moves away from the center.
\par Let us now estimate the order of magnitude of the radial velocity of
gas(and stars) in the galactic disk.  The radius of our galaxy is $50000$ light
years and the age of our galaxy is of the order of magnitude of the age of the
Universe, $10^{17}$s. The gas which is at the edge of the disk must have
moved from the center with an average velocity of 5km/s. This is just a rough
estimation.  It is very important to obtain the
mean radial velocities of the spiral arms of the Milky Way to compare with
the above figure.
\vskip .15in
\noindent
XIV. THE MAGNITUDE OF THE SUPERWEAK FORCE
\vskip .25in
\par We may estimate the strength of the superweak force in the following way.
A star(or the material that formed it) near the edge of our galaxy's disk has
traveled 50,000 light years in about $10^{17}$s. Therefore, the order of
magnitude of its acceleration is about $a=10^{-13}$ms$^{-2}$. This figure is
the upper limit of the accuracy of the experiments which have been performed so
far on the fifth force$^{(23)}$. Let us, now, have an estimation on the values
of $(A_B)^{1/2}g$ and $(A_I)^{1/2}g$. As has been previously discussed, the
above star(or the material that formed it) has left the center of our galaxy
about 10 billion years ago. The superweak force which acted upon the star when
it left the center was of the order of
\begin{eqnarray} \frac{\left(\sqrt{A_B} - \sqrt{A_I}\right)^{2}g
{2}Bb}{r^2}
&=& bm_{p}a\end{eqnarray}
\noindent

where $B$ and $b$ are the number of baryons of the galaxy and of the star,
respectively; $m_p$ is the proton's mass; and $r$ is the distance between the
star and the center of the galaxy. This distance was very small, probably a few
light years. Let us take for $r$ the size of the center of our galaxy, about
$40$ light years$^{(39)}$. Therefore, the order of magnitude of $\sqrt{A_B}g$
and $\sqrt{A_I}g$ is $10^{-37}$(in MKSC units). We, clearly, see that the
coupling constant of the superweak interaction is extremely small. The
superweak force between two 1kg masses separated by 1m is only $10^{-20}$
Newton, which is $10^{10}$ times smaller than the gravitational force between
the two masses.
\vskip .25in
\noindent
XV. VARIATION OF THE GRAVITATIONAL CONSTANT
\vskip .15in
\par In the light of the present theory we can discuss Dirac's Large Number
Hypothesis$^{48}$. Dirac considered the ratio of the electrostatic  force
to the gravitational force between the electron and the proton in the hydrogen
atom, which is given by
\begin{eqnarray}
& &\frac{e^2}{Gm_{e}m_{p}}{\approx}2\times{10^{39}}
\end{eqnarray}

\noindent
where $e$ is the charge of the electron(and pronton), $G$ is the gravitational
constant, and $m_e$, $m_p$ are the masses of the electron and proton
respectively. How can we understand the meaning of this large number? Taking
a look at Tables 1 and 2 we can understand the reason of this number. Galaxies
were formed out of hydrogen atoms(mainly), i.e., the gravitational force was
only able to form something stable {\it{by means}} of the build up of
atoms(electromagnetic force). Therefore, this number just represents a relation
between these two bound states of matter. Of course, we have other numbers
relating the other bound states(of the universe), i.e, relating any two
fundamental forces.
\par
Dirac also expressed the age of the Universe in terms of the
time that light takes to traverse a classical electron. Taking this time as
a basic unit the present
age of the Universe(18 billion years) may be expressed as
\begin{eqnarray}
t{\approx}7\times{10^{39}}\frac{e^2}{m_{e}c^3}.
\end{eqnarray}

He concluded, that, since the age of the universe is increasing with time,
$G$ must vary inversely proportionally to time. This could be true only if
the gravitational force were the ultimate force for the universe.  Since
this is not the case, $G$ may not vary with time.  As was shown above the
age of the universe depends(mainly) on the strength of the superweak force.

\vskip .25in
\noindent
XV. PRELIMINARY IDEAS ON PREQUARKS
\vskip .15in
\par The classification of matter achieved in the beginning of the paper
implied that quarks are formed of prequarks. Let us develop some preliminary
ideas which may help us towards the understanding of the superstrong
interaction.  This interaction may  manifest itself as corrections to the
interaction among quarks.  Presumably, just as quarks do, prequarks are
supposed to be permanently confined inside baryons.
\par Actually, the composition of quarks is an old idea, although it has been
proposed on different grounds$
{(49)}$. A major distinction is that in this
paper leptons are elementary particles. This is actually consistent with the
smallness of the electron mass which is already too small for a particle with
a very small radius$^{(50)}$.
\par In order to distinguish the model proposed in this paper from several
other models of the literature we will name these prequarks with a different
name.   We may call them {\it{primons}}, a word derived from the latin word
{\it{primus}} which means first.
\par From the above considerations quarks are composed of primons. Also, we
saw above that there must be a sort of polarization in the formation of a
quark.  The best candidate is the exchange of a particle which
may be the carrier of the superstrong force.  Since a baryon is composed of
three quarks, it is reasonable to consider that a quark is composed of two
primons which are polarized by the exchange of some
kind of charge which is carried on by the corresponding boson.
Let us name the exchanged particle the mixon(this choice will become
clearer below).
\par In order to reproduce the spectrum of 18 quarks(6 quarks in 3 color
states) we need 12 primons(4 primons in 3 supercolor states).  Therefore, we
have 4 triplets. As to the charge, one triplet has charge (5/6)e and any other
triplet has charge (-1/6)e.
\par Let us assume that the exchanged particle is a boson. Since quarks have
spin $1/2$, the spin of each primon has  to be equal to 1/4. In this case the
boson has spin zero. At this stage we may introduce a postulate concerning the
unit of quantization. As is well known the unit of quantization is $\hbar$.
This unit of measure is arbitrary and was taken as such in order to have an
agreement between experimental data and theory at the level of atomic physics.
Later on this unit of measure was applied to elementary particles and up to
the level of quarks it still holds.  However, at the level of the quarks
constituents it may not hold anymore.  We may postulate that at the level of
primons the unit of measure is $\bar{H}=\hbar/2$. In this way primons are also
fermions with a spin given by
\begin{eqnarray} s&=&\left(\frac{1}{2}\right)\bar{H}.
\end{eqnarray}

As we will see below, this is consistent with the required properties which are
needed for forming quarks with primons.
\par Let us try to arrive at a possible equation for free primons beginning
with Dirac's equation,
\begin{eqnarray} \left(i{\hbar}{\alpha}_{\mu}\frac{\partial}{\partial{x}^{\mu}}
- {\beta}mc\right)\Psi &=& 0 \end{eqnarray}

\noindent
where  $\mu=0,1,2,3$.
\noindent
If, now, we divide this equation by 2, and make the substitutions
${\hbar}/2=\bar{H}$ and ${\bar{\beta}}/2=\beta$, we obtain of course another
Dirac equation.  By imposing that each component $\Psi_{\sigma}$ of $\Psi$
must satisfy Klein-Gordon equation we obtain the following algebra, which is
slightly different from Dirac's,
\begin{eqnarray*}\alpha_{i}\alpha_{k} + \alpha_{k}\alpha_{i} &=& 2\delta_{ik}
\nonumber \\
                \alpha_{i}\bar{\beta} + \bar{\beta}\alpha_{i} &=& 0\nonumber \\
                {\alpha_{i}}^{2} &=& 1 \nonumber \\
                {\bar{\beta}}^{2} &=& \frac{1}{4}.
\end{eqnarray*}

Of course, $\alpha_{i}$ are the same as in Dirac's equation, but now
$\bar{\beta}={\beta}/2$.
\par The new Dirac equation becomes
\begin{eqnarray} \left(i\bar{H}\alpha_{\mu}\frac{\partial}{\partial{x}^{\mu}} -
{\bar{\beta}}mc\right)\Psi &=&0 \end{eqnarray}

\noindent
where $\bar{H}$ is a new unit of quantization and $\bar{\beta}$ is Dirac's
${\beta}/2$. Of course, this equation is also Lorentz covariant.
\par Since quarks have spin equal to 1/2, only primons with parallel(or
antiparallel) spins form bound states(quarks). This means that the spin
wave function of a bound state(quark) is symmetric and, because the total wave
function is antisymmetric, the rest of the wave function(which includes the
superflavor, supercolor and spatial parts) has to be antisymmetric.
\par The superstrong interaction is such that only primons with different
quantum numbers form bound states, that is, quarks. Taking into account the
above considerations on spin and charge, we have the following
table for primons(Table III).  With this table we are able to form all quarks
as is shown in Table IV. The colors are formed from the supercolors as shown
in Table V.
\par Therefore, we assume that a prequark transforms $p_{ij}$, with
\begin{itemize}
\item superflavor index: $i=1,2,3,4$
\item supercolor index: $j={\alpha},{\beta},{\gamma}$.
\end{itemize}

These indices must transform respectively under $SU(4)|_{superflavor}$ and
$SU(3)|_{supercolor}$.  Because of the selection rules the group
$SU(3)|_{supercolor}$ is reduced to the subgroup $SU(2)|$, and therefore the
number of bosons must be 3.
\par As was said at the beginning of this section the  ideas on prequarks
are very preliminary and a deeper understanding of the
superstrong interaction, as proposed in this paper, is under consideration.
This understanding must begin defining the internal quantum numbers for
the interaction, i.e., substitutes for hypercharge, isospin, and so on.
\vskip .25in
\noindent
XVI. THE FUNDAMENTAL INTERACTIONS OF MATTER
\vskip .2in
\hskip 3.5in {\it{A maximis ad minima}}
\vskip .15in
\par It is well known that the electromagnetic interaction is mediated by a
massless boson, the photon.  The weak interaction is mediated by the three
heavy vector bosons, $W^{+},W^{-}$ and $Z^{o}$. It was shown by Weinberg, Salam
and Glashow that the weak and electromagnetic interactions are unified at short
distances. In the domain of the atomic nucleus the strong interaction is
mediated by the three pions, $\pi^{+},\pi^{-}$ and $\pi^{o}$. It has been shown
in this paper that the strong and superweak interactions are unified at $t=0$.
This means that total unification may not be possible and that the fundamental
interactions are unified in pairs. In this paper it was assumed that the range
of the superweak interaction is infinite. This means that its mediator is a
massless boson which, of course, must be linked to baryon number conservation
and isospin. Let us name this boson as  the symmetron.
\par We have two interactions left, the superstrong and the gravitational
interactions. We may suppose that they are also unified at $t=0$.  As was shown
in the previous section that there must exist 3 bosons mediating the
superstrong interaction.  The gravitational interaction is presumably mediated
by the graviton, which is a massless boson of spin 2. The data for all
interactions of nature are shown in Table VI.
\par As is proposed in this paper the forces of nature are unified in pairs.
It is interesting to notice that we have the table below:
\vskip .2in
\centerline{UNIFICATION OF THE FORCES OF NATURE}
\vskip .25in
\centerline{Weak(3 bosons) with Electromagnetic(1 boson)}
\centerline{Strong(3 bosons) with Superweak(1 boson)}
\centerline{Superstrong(3 bosons) with Gravity(1 boson) ???}
\vskip .25in
\noindent
XVII. QUARK CONFINEMENT
\vskip .15in
\par Present experiments show that quarks are permanently confined inside
hadrons, at least, within the energies allowed by the present generation of
particle accelerators.
\par Quark confinement can be explained as the result of the superstrong
interaction. Because the quark belongs to a polarized state it is formed by
the action of the superstrong and strong forces. That is, we expect that there
must exist an effective attractive potential  as we have in the other kinds of
polarized states. Therefore, we expect to have a sort of Lennard-Jones
effective potential in the interaction among quarks. Expanding a
Lennard-Jones potential around the minimum we obtain a harmonic oscillator
potential.  Thus, if we consider that in their lowest state of energy quarks
are separated by a distance $r_{q}$, then for small departures from equilibrium
the potential must be of the form
\begin{eqnarray} V(r) &=& V_{o} + K(r - r_{q})^{2}\end{eqnarray}
\noindent

where $K$ is a constant and $V_{o}$ is a negative constant representing the
deepness of the potential well. As $r$ increases
the restoring force among quarks also increases. Because of it quarks may be
permanently confined.
\par Quark confinement is in agreement with other considerations and results
of this article, for as was shown above the superweak force must exist at
$t=0$ in order to make the expansion begin. {\it{This may only happen if
neutrons and protons are present at $t=0$}}. This, in turn, means that quarks
are not free at $t=0$. That is, it means that they are confined inside the
nucleons.  We will see more on quark confinement below, when we will find
out the behavior of the quark wavefunction with the separation $r$ between
two quarks.
\vskip .25in
\noindent
XVIII. THE ENERGIES OF BARYON STATES(INCLUDING RESONANCES)
\vskip .15in
\par We have deduced above that there must exist a sort of Lennard-Jones
potential in the interaction among quarks. Around its minimum we may
approximate this potential by the potential of a harmonic oscillator and
include the anharmonicity as a perturbation. By doing so we may be able to
calculate the energies of almost all baryon states.
\par Let us consider a system composed of three quarks which interact in pairs
by means of a harmonic potential. Let us disregard the electromagnetic
interaction which must be considered as a perturbation. Also, let us  disregard
any rotational contribution which must enter as  perturbation too. This is
reasonable because the strong and superwtrong interactions must be much larger
than the ``centrifugal'' potential. If we disregard the spin interaction
among quarks, we may just use Schr\"{o}dinger equation in terms of normal
coordinates$^{51}$
\begin{eqnarray} \sum_{i=1}^{6}
\frac{{\partial}^{2}\psi}{{\partial}{\xi}_{i}^{2}} +
\frac{2}{{\hbar}^2}\left(E -
\frac{1}{2}\sum_{i=1}^{6}{\omega}_{i}{{\xi}^2}\right)\psi &=& 0\end{eqnarray}

\noindent
where we have used the fact that the three quarks are always in a plane. The
above equation may be resolved into a sum of 6 equations
\begin{eqnarray} \frac{{\partial}^{2}\psi}{{\partial}{\xi}_{i}^{2}} +
\frac{2}{{\hbar}^2}\left(E_{i} - \frac{1}{2}\omega_{i}\xi^{2}\right) &=& 0,
\end{eqnarray}
\noindent
which is the equation of a single harmonic oscillator of potential energy
$\frac{1}{2}\omega_{i}\xi^{2}$ and unitary mass with
\begin{eqnarray} E &=& \sum_{i=1}^{6} E_{i}.\end{eqnarray}

\noindent
The general solution is a superposition of 6 harmonic motions in the 6 normal
coordinates.
\par The eigenfunctions $\psi_{i}(\xi)$ are the ordinary harmonic oscillator
eigenfuntions
\begin{eqnarray} \psi_{i}(\xi_{i}) &=& N_{v_i}e^{(\alpha_{i}/2)\xi_{i}^{2}}
H_{v_i}(\sqrt{\alpha_{i}}\xi_{i}),\end{eqnarray}

\noindent
where $N_{v_i}$ is a normalization constant, $\alpha_{i} = \nu_{i}/{\hbar}$ and
$H_{v_i}(\sqrt{\alpha_{i}}\xi_{i})$ is a Hermite polynomial of the $v_i$th
degree. For large $\xi_{i}$ the eigenfunctions are governed by the exponential
functions which make the eigenfunctions go to zero very fast.  Of course,
{\it{this is valid for any energy and must be the reason behind quark
confinement}}. We will come back yet to this point after calculating the
possible energy levels of the baryons.
\par The energy of each harmonic oscillator is
\begin{eqnarray} E_{i} &=& h\nu_{i}(v_{i} + \frac{1}{2}),\end{eqnarray}

\noindent
where $v_{i} = 0,1,2,3,...$ and $\nu_i$ is the classical oscillation frequency
of the normal vibration $i$, and $v_i$ is the vibrational quantum number.
The total vibrational energy of the system can assume only the values
\begin{eqnarray} E(v_{1},v_{2},v_{3}, ...v_{6}) &=& h\nu_{1}(v_{1} +
\frac{1}{2}) + h\nu_{2}(v_{2} + \frac{1}{2}) + ... h\nu_{6}(v_{6} +
\frac{1}{2}.\end{eqnarray}

\par The three quarks in a baryon must always be in a plane. Therefore, each
quark is composed of two oscillators and so we may rearrange the energy
expression as
\begin{eqnarray} E(n,m,k) &=& h\nu_{1}(n + 1) + h\nu_{2}(m + 1) +
h\nu_{3}(k + 1),\end{eqnarray}

\noindent
where $n=v_{1} + v_{2},m=v_{3} + v_{4},k=v_{5} + v_{6}$. Of course, $n,m,k$ can
assume the values, 0,1,2,3,... We may find the constants $h\nu$ from the ground
states of some baryons. They are the known quark masses taken as$^{53}$
$m_{u}=m_{d}= .31$Gev, $m_{s}= .5$Gev, $m_{c}=1.5$Gev and $m_{b}=5$Gev. The
mass of the top quark has not been determined, but as we will show the present
theory may help in the search for its mass.
\par Let us start the calculation with the states ddu(neutron), uud(proton) and
ddd($\Delta^{-}$), uuu($\Delta^{++}$) and their resonances. Because
$m_{u}=m_{d}$, we have that their energies must be given by(in Gev)
\begin{eqnarray} E_{n,m,k} &=& .31(n+m+k + 3).\end{eqnarray}

\noindent
The calculated values are displayed in Table 7. The agreement between the
calculated values and the experimental data is quite good
except for a very small number of states. We will discuss this later on.
\par The energies of the particles $\Sigma$ and $\Lambda$ and their resonances
are given by(in Gev)
\begin{eqnarray} E_{n,m,k} &=& .31(n+m+2) + .5(k+1).\end{eqnarray}

\noindent
Again the agreement with the experimental data is very good except for a
few states(Table 8).
\par For the $\Xi$ particle the energies are expressed by(in Gev)
\begin{eqnarray} E_{n,m,k} &=& .31(n+1) + .5(m+k+2).\end{eqnarray}
\noindent
See Table 9 to check the agreement with the experimental data.
\par In the same way for the particles $\Omega$ and $\Delta_{c} {+}$ we
have the following formulas, respectively:
\begin{eqnarray} E_{n,m,k} &=& .5(n+m+k+3),\end{eqnarray}
\noindent
and
\begin{eqnarray} E_{n,m,k} &=& .31(n+m+2) + 1.5(k+1).\end{eqnarray}
\par In the same fashion we list below the formulas for calculating the
energies of many other states:
\begin{itemize}
\item ucc, $E_{n,m,k} = .31(n+1) + 1.5(m+k+2)$;
\item ssc, $E_{n,m,k} = .5(n+m+2) + 1.5(k+1)$;
\item scc, $E_{n,m,k} = .5(n+1) + 1.5(m+k+2)$;
\item ccc, $E_{n,m,k} = 1.5(n+m+k+3)$;
\item ccb, $E_{n,m,k} = 1.5(n+m+2) +5(k+1)$;
\item cbb, $E_{n,m,k} = 1.5(n+1) + 5(m+k+2)$;
\item ubb and dbb, $E_{n,m,k} = .31(n+1) + 5(m+k+2)$
\item uub and ddb, $E_{n,m,k} = .31(n+m+2) + 5(k+1)$
\item bbb, $E_{n,m,k} = 5(n+m+k+3)$.
\end{itemize}

\par We clearly see from Tables 7, 8 and 9 that there are discrepancies in some
states between the calculated energy and the experimental data. This is
expected because of the assumptions that we made. An improved model must
must include the electromagnetic interaction and the effect of rotation(spin).
Both will split the levels. Also, we must include the anharmonicity of the
potential. The present treatment assumed that the oscillators are independent.
But, this may not be the case and there must exist coupling between them. We
clearly see that the overtones coincide with the main levels.
\par The states $\Lambda$(1.520), $N$(1.670, 1.688, 1.700, 1.780),
$\Delta$(1.650, 1.670)  were not reproduced but they can be built up with
transitions from other states. The decay of $\Lambda$(1.520) shows that it may
be constructed from the ${\Sigma}(1.385) + \pi$, that is, $1.520 {\sim} 1.385
+ .140$. This is not just a pure summation of two numbers. It represents the
transition from the level $\Sigma$(1.385) plus the energy of the boson, in
the same manner as we have with electronic transitions in an atom. In the
same fashion $N$(1.670), $N$(1.688) and $N$(1.700) are combinations of
$N(1.535) + \pi$, and $N$(1.780) may be formed from $N$(1.535) by means of
a two pion transition. The states $\Delta$(1.650, 1.670) may be formed from
the state $N$(1.535) plus a pion transition.
\par From the above discussion we see clearly that it is meaningless to
try to make quarks free by using more energetic collisions  between
two baryons because we will just get more excited states. That is, we will
obtain more and more resonances. This is so because, as as shown above,
as the separation between two quarks increases the wavefunction vanishes very
fast regardless of the state of energy.
\par The parity of the spatial part of the wavefunction is given by
$(-1)^{m+n+k}$. The multiplicity of each level must be given by $2I +1$, where
$I$ is the isospin.

\vskip .25in
\noindent
XIX. CONCLUSION
\vskip .15in
\par The present theory rules out any role of the exotic dark matters
which  have been proposed in the literature as serving as the missing mass
necessary for closing the universe. It is shown a way of having a closed
universe by means of another force which brings the universe back to its
`beginning' even having ${\Omega}<1$.
\par It is shown that the universal expansion must be slowing down at the
present epoch and that the expansion is not so fast around $t=0$. It is just
an expansion rather than an explosion.
\par A possible galactic evolution which is consistent with  the
observational data has been presented. It is given a reasonable explanation
for the flat rotational curve($v{\times}r$) of spiral galaxies.  It is shown
that the flat rotational curve of spirals is directly connected with the
spiral structure itself and with the evolution of such galaxies.
\par It is proposed that nature has six fundamental forces which are unified in
pairs and, therefore, reduced to three at $t=0$. Some general ideas concerning
the characteristics of the superstrong interaction have been presented.
\par A reasonable physical explanation has been provided for quark confinement.
The energies of most baryon states are calculated in a simple manner.
\par It is shown that it is consistent to have a contracting Universe  of
antimatter.
\par It is expected that the `galactic liquid'(with its superstructures) is
quite complex just as all normal liquids are.  It has to be described by the
superweak and gravitational forces. We intend to further the studies in this
direction towards understanding its formation and structure.
\par Taking into account the existence of this new interaction we must
investigate the formation of the light elements and rethink concepts such
as the entropy of the universe and its apparent order.
\par It is hoped that in the near future experimenters will find evidence
of the superweak force.
\vskip .25in
\noindent
\large{Acknowledgments}
\vskip .15in
\normalsize
\par
I would like to thank I.K.Sou and X.Chu from the University of
Illinois at Chicago,  P.Panigrahi from the University of Ottawa, and Claudio
Macedo and Lafaete Bezerra from the Universidade Federal de Sergipe for
valuable discussions  on the paper. I also thank P.S.Wijewarnasuriya, Julie
\v{C}iapas, and Romualdo Lima Silva  for their technical assistance.
\newpage

\vskip .4in
\Large{References}
\normalsize
\vskip .3in
\noindent
1. A. A. Penzias and R. W. Wilson, {\it{Astrophys. J.}} {\bf{142}}, 419(1965).
\newline
2. J. C. Mather, Cheng,E.S., Eplee,R.E., Isaacman,R.B., Meyer,S.S.,
Shafer,R.A., Weiss,R., Wright,E.L., Bennet,C.L., Boggess,N.W., Dwek,E.,
Gulkis,S., Hauser,M.G., Janssen,M., Kelsall,T., Lubin,P.M., Moseley,Jr.,S.H.,
 Murdock,T.L., Siverberger,R.F., Smoot,G.F.,and Wilkinson,D.T., {\it{Astrophys.
J.(Letters)}} {\bf{354}}, L37(1990).
\newline
3. E. Hubble, in {\it{Proceedings of the National Academy of Science}},
{\bf{15}},168-173(1929).
\newline
4. J. C. Mather, talk at the XIII Interantional Conference on
General Relativity and Gravitation, Huerta Grande, Argentina, 1992.
\newline
5. V. de Lapparent, M. J. Geller, and J. P. Huchra, {\it{Astrophys.
J.(Letters)}} {\bf{302}}, L1(1986).
\newline
6. T. J. Broadhurst, R. S. Ellis, D. C. Koo, and A. S. Szalay, {\it{Nature}}
{\bf{343}},726(1990).
\newline
7. A. Dressler and S. M. Faber, {\it{Astrophys. J.}} {\bf{354}}, 13(1990).
\newline
8. A. Dressler, S. M. Faber, D. Burnstein, R. L. Davies, D. Lynden-Bell,
R. J. Terlevich, and G. Wegner, {\it{Astrophys. J.(Letters)}} {\bf{313}},
L37(19
\newline
9. D. Lynden-Bell, S. M. Faber, D. Burnstein, R. L. Davies, A. Dressler,
R. J. Terlevich, and G. Wegner, {\it{Astrophys. J.}} {\bf{326}}, 19(1988).
\newline
10. H. Kurki-Suonio, G. J. Mathews, and G. M. Fuller, {\it{Astrophys.
J.(Letters)}} {\bf{356}}, L5(1990).
\newline
11. S. J. Maddox, G. Efstathiou, W. J. Sutherland and J. Loveday, {\it{Mon.
Not.Astr.Soc.}} {\bf{242}}, 43p(1990).
\newline
12. W. Saunders. M. Rowan-Robinson, and A. Lawrence, ``The Spatial Correlation
Function of IRAS Galaxies on Small and Intermediate Scales'', 1992, QMW
preprint.
\newline
13. G. B. Dalton, G. Efstathiou, S. J. Maddox, and W. J. Sutherland,
{\it{Astrophys. J. Lett.}} {\bf{390}}, L1, 1992.
\newline
14. M. S. Vogeley, C. Park, M. J. Geller, and J. P. Huchra, {\it{Astrophys.
J. Lett.}} {\bf{391}}, L5, 1992.
\newline
15. C. Park, J. R. Gott, and L. N. da Costa, {\it{Astrophys.  J. Lett.}}
{\bf{392}}, L51, 1992.
\newline
16. M. Rowan-Robinson, {\it{New Scientist}} {\bf{1759}}, 30(1991).
\newline
17. W. Saunders, C. Frenk, M. Rowan-Robinson, G. Efstathiou, A. Lawrence,
N. Kaiser, R. Ellis, J. Crawford, X.-Yang Xia and I. Parry, {\it{Nature}}
{\bf{349}}, 32(1991).
\newline
18. S. White, talk at the XIII International Conference on General
Relativity and Gravitation, Huerta Grande, Argentina, 1992.
\newline
19. H. Fritzsch, in {\it{Proceedings of the twenty-second Course of the
International School of Subnuclear Physics, 1984}}, ed. by A. Zichichi
(Plenum Press, New York, 1988).
\newline
20. W. K\"{u}pper, G. Wegmann, and E. R. Hilf, {\it{Ann. Phys.}} {\bf{88}},
454(
\newline
21. D. Bandyopadhyay, J. N. De, S. K. Samaddar, and D. Sperber, {\it{Phys.
Lett. B}} {\bf{218}}, 391.
\newline
22. E. Fischbach, in {\it{Proceedings of the NATO Advanced Study Institute on
Gravitational Measurements, Fundamental Metrology and Constants, 1987}},
ed. by V. de Sabbata and V. N. Melnikov(D. Reidel Publishing Company,
Dordrecht,
Holland, 1988).
\newline
23. E. G. Adelberger, B. R. Heckel, C. W. Stubbs and W. F. Rogers, {\it{Annu.
Rev. Nucl. Part. Sci.}} {\bf{41}}, 269(1991).
\newline
24. C. M. Will, talk at the XIII International Conference on General
Relativity and Gravitation, Huerta Grande, Argentina, 1992.
\newline
25. S. L. Shapiro and S. A. Teukolsky, {\it{Black Holes, White Dwarfs and
Neutron Stars}}(Wiley, New York, 1983).
\newline
26. J. H. Oort, {it{Bull. Astron. Inst. Neth.}} {\bf{15}}, 45(1960).
\newline
27. R. A. Alpher, J.W.Follin,Jr., and R.C.Herman, {\it{Phys.Rev.}}
{\bf{92}},1347(1953).
\newline
28. G. Gamow, {\it{Phys. Rev.}} {\bf{70}}, 572(1946).
\newline
29. R. A. Alpher and R.C.Herman, {\it{Revs. Modern Phys.}} {\bf{22}},
153(1950).
\newline
30. M. Rees, {\it{Phys. Rev. Lett.}} {\bf{28}},1669(1972).
\newline
31. D. Layzer and R. M. Hively, {\it{Astrophys. J.}} {\bf{179}},361(1973).
\newline
32. M. Rees, {\it{Nature}} {\bf{275}}, 35(1978).
\newline
33. B. J. Carr, {\it{Mon. Not. R. Astr. Soc.}} {\bf{181}}, 293(1977).
\newline
34. B. J. Carr, {\it{Mon. Not. R. Astr. Soc.}} {\bf{195}}, 669(1981).
\newline
35. B. J. Carr, J. R. Bond and W. D. Arnett, {\it{Astrophys. J.}} {\bf{277}},
445(1984).
\newline
36. M. V. Berry, in {\it{Principles of Cosmology and Gravitation}}(Adam Hilger,
Bristol, 1991).
\newline
37. K. L. Visnovsky, C. D. Impey, C. B. Foltz, P. C. Hewett, R. J. Weymann and
S. L. Morris, {\it{Astrophys. J.}} {\bf{391}}, 560(1992).
\newline
38. T. A. Boroson and S. E. Persson, {\it{Astrophys. J.}} {\bf{293}},
120(1985).
\newline
39. W. J. Kaufmann,III, in {\it{Galaxies and Quasars}}(W.H.Freeman and Company,
San Francisco, 1979).
\newline
40. T. M. Heckman, K. C. Chambers and M. Postman, {\it{Astrophys. J.}},
{\bf{391}}, 39(1992).
\newline
41. S. Baum and T. M. Heckman, {\it{Astrophys. J}} {\bf{336}}, 702(1989).
\newline
42. N. Jackson and I. Browne, {\it{Nature}}, {\bf{343}}, 43(1990).
\newline
43. A. Lawrence, {\it{Mon. Not. R. Astr. Soc.}}, 1992, in press.
\newline
44. R. W. Goodrich and M. H. Cohen, {\it{Astrophys. J.}} {\it{391}}, 623(1992).
\newline
45. Y. Sofue, {\it{Astro. Lett. Comm.}} {\bf{28}}, 1(1990).
\newline
46. N. Nakai, M. Hayashi, T. Handa, Y. Sofue, T. Hasegawa and M. Sasaki,
{\it{Pub. Astr. Soc. Japan}}{\bf{39}}, 685(1987).
\newline
47. C. A. Beichman, {\it{Astro. Lett. Comm.}} {\bf{27}}, 67(1988).
\newline
48. P. A. M. Dirac, in {\it{Directions in Physics}} (Wiley, New York, 1978).
\newline
49. K. Huang, in {\it{Quarks, Leptons and Gauge Fields}}(World Scientific,
Singapore, 1982).
\newline
50. G. 'tHooft, in {\it{Recent Developments in Gauge Theories}}, eds. G.
'tHooft et al.(Plenum Press, New York, 1980).
\newline
51. L. Pauling and E. B. Wilson Jr., {\it{Introduction to Quantum Mechanics}}
(McGraw-Hill, New York, 1935).
\newline
52. L. Ryder, {\it{Elementary Particles and Symmetries}} (Gordon Breach Science
Publishers, New York, 1975).
\newpage
\rule{0in}{3in}
\large
ATTENTION!  ATTENTION!
\vskip .2in

Table 6 has been separated into two tables. You will have to join them.
\newpage
\normalsize
\begin{center}
\begin{tabular}{c c c c c c c} \hline\hline\\
& ? & & quark & & nucleon & \\
\\
& nucleon & & nucleus & & atom & \\
\\
& atom & & gas & & galaxy & \\
& & & liquid & & & \\
& & & solid & & & \\
\\
& galaxy & & galactic liquid & & ? \\
\\
\hline\hline\\
\end{tabular}
\end{center}
\vskip .5in

\noindent
\centerline{Table 1. The two general states  which make everything in the
Universe,}
\centerline{stepwise. The table is arranged in such a way as to show the links
between}
\centerline{the polarized states and the single states.}

\newpage
\rule{0in}{1.5in}
\begin{center}
\begin{tabular}{c c c} \hline\hline\\ ? & ? & strong force \\
& strong force & \\
\\
strong force & strong force & electromagnetic force \\
& electromagnetic force & \\
\\
electromagnetic force & electromagnetic force & gravitational force \\
& gravitational force & \\
\\
gravitational force & gravitational force & ?\\
& ? & \\
\\
\hline\hline\\
\end{tabular}
\end{center}
\vskip .5in
\noindent
\centerline{Table 2. Three of the fundamental forces of nature. Each force
appears}
\centerline{three times and is linked to another force by means of a polarized
state.}
\centerline{Compare with Table I}.

\newpage
\begin{center}
\begin{tabular}{  cc  ccccc } \hline\hline
& & & & & & \\
superflavor & & & charge & & spin & \\
& & & & & & \\
\hline \hline
& & & & & & \\
$p_{1}$ & & & $\frac{5}{6}$  & & $\frac{1}{2}$ & \\
& & & & & & \\
\hline
& & & & & & \\
$p_{2}$ & & & $-\frac{1}{6}$ & & $\frac{1}{2}$ & \\
& & & & & & \\
\hline
& & & & & & \\
$p_{3}$ & & & $-\frac{1}{6}$ & & $\frac{1}{2}$ & \\
& & & & & & \\
\hline
& & & & & & \\
$p_{4}$ & & & $-\frac{1}{6}$ & & $\frac{1}{2}$ & \\
& & & & & & \\
\hline\hline
\end{tabular}
\end{center}
\vskip .5in

\noindent
\centerline{Table 3. Table of charges and spins of primons.}

\newpage
\begin{center}
\begin{tabular}{  cc  ccccccccccccc } \hline\hline
& & & & & & & & & & & & & & \\
 & & & $p_{1}$ & & & $p_{2}$ & & & $p_{3}$ & & & $p_{4}$ & & \\
& & & & & & & & & & & & & & \\
\hline\hline
& & & & & & & & & & & & & & \\
$p_{1}$ & & &   & & & u & & & s & & & t & & \\
& & & & & & & & & & & & & & \\
\hline
& & & & & & & & & & & & & & \\
$p_{2}$ & & & u & & &   & & & d & & & c & & \\
& & & & & & & & & & & & & & \\
\hline
& & & & & & & & & & & & & & \\
$p_{3}$ & & & s & & & d & & &   & & & b & & \\
& & & & & & & & & & & & & & \\
\hline
& & & & & & & & & & & & & & \\
$p_{4}$ & & & t & & & c & & & b & & &   & & \\
& & & & & & & & & & & & & & \\
\hline\hline
\end{tabular}
\end{center}
\vskip .5in

\noindent
\centerline{Table 4. Table of composition of quark flavors.}

\newpage
\begin{center}
\begin{tabular}{ cc  ccccccc } \hline\hline
& & & & & & & & \\
         & & & $\alpha$ & & $\beta$ & & $\gamma$ & \\
& & & & & & & & \\
\hline\hline
& & & & & & & & \\
$\alpha$ & & &          & & blue   & & green    & \\
& & & & & & & & \\
\hline
& & & & & & & & \\
$\beta$  & & &  blue    & &        & & red      & \\
& & & & & & & & \\
\hline
& & & & & & & & \\
$\gamma$ & & & green    & & red    & &          & \\
& & & & & & & & \\
\hline\hline

\end{tabular}
\end{center}
\vskip .5in

\noindent
\centerline{Table 5. Table of the generation of colors}
\centerline{out of the supercolors.}

\newpage
\begin{center}
\begin{tabular}{ c  c c c c c } \hline
& & & & & \\
Interaction & Superstrong & Strong & Electromagnetic & Gravity & Superweak \\
& & & & & \\
\hline\hline
& & & & & \\
Static  & $\frac{{\mho}_{1}{\mho}_{2}}{r}e^{-{\mu}r}$? &
-$\frac{\sqrt{g}\sqrt{g}}{4{\pi}r}e^{-{\mu}r}$
& $\frac{q_{1}q_{2}}{4{\pi}r}e^{-{\mu}r}$
& - $\gamma\frac{m_{1}m_{2}}{r}e^{-{\mu}r}$
& $\frac{Q_{1}Q_{2}}{r}e^{-{\mu}r}$
\\
potential & $\frac{\hbar}{{\mu}c}{l}10^{-13}$
& $\frac{\hbar}{{\mu}c}{\sim}10^{-13}$
& ${\mu}={\infty}$ & ${\mu}={\infty}$ & $
{\mu}={\infty}$ \\
& & & & & \\
\hline
& & & & & \\
Coupling & ?  & $\frac{g^2}{4{\pi}{\hbar}c}{\approx}10$
& $\frac{e^2}{4{\pi}{\hbar}c}=\frac{1}{137.036}$
& $\frac{{\gamma}{m_{p}}^{2}}{{\hbar}c}=5.76{\times}10^{-36}$
& $A_{B}, A_{I} {\approx} 10^{-67}$ \\
& & & & $m_p$ = proton mass & \\
& & & & & \\
\hline
& & & & & \\
Bosons & 3 mixons & ${\pi}^{+},{\pi}^{-},{\pi}^{0}$
& photon & graviton & symmetron \\
& & & & &  \\
\hline\hline
\end{tabular}
\end{center}
\vskip .6in

\noindent
\centerline{Table IV. The Six Interactions of Nature}

\newpage
\begin{center}
\begin{tabular}{ c c } \hline
& \\
Superweak & Weak \\
& \\
\hline\hline
& \\
$\frac{Q_{1}Q_{2}}{r}e^{-{\mu}r}$ & none \\
${\mu}={\infty}$ & \\
& \\
\hline
& \\
${A_{B}, A_{I}} {\approx} 10^{-67}$  & ${\Lambda}{m_p}^{2}
= 1.01{\times}10^{-5}
$  \\
& \\
& \\
\hline
& \\
symmetron & $W^{+},W^{-},Z^{0}$ \\
&  \\
\hline\hline
\end{tabular}
\end{center}
\vskip .6in

\noindent
\centerline{Table IV. The Six Interactions of Nature}

\newpage
\begin{center}
\begin{tabular}{ c  l l l c c } \hline
& & & & & \\
$n,m,k$ & $E_{C}(Gev)$ & $E_{M}$(Gev) & Error{\%} & Isospin & Parity\\
& & & & & \\
\hline\hline
& & & & & \\
0,0,0  & 0.93  & 0.938($N$) & 0.96 & 1/2 & even \\
& & & & & \\
\hline
& & & & & \\
$m+n+k=1$ & 1.24 & 1.24($\Delta$) & 0 & 3/2 & odd \\
& & & & & \\
\hline
& & & & & \\
$m+n+k=2$ & 1.55 & 1.47($N$) & 5.2 & 1/2 & even \\
$m+n+k=2$ & 1.55 & 1.52($N$) & 1.9 & 1/2 & even \\
$m+n+k=2$ & 1.55 & 1.535($N$) & 1.0 & 1/2 & even \\
& & & & & \\
\hline
& & & & & \\
$m+n+k=3$ & 1.86 & 1.86($N$) & 0 & 1/2 & odd \\
$m+n+k=3$ & 1.86 & 1.89($\Delta$) & 1.6 & 3/2 & odd \\
$m+n+k=3$ & 1.86 & 1.91($\Delta$) & 2.7 & 3/2 & odd \\
$m+n+k=3$ & 1.86 & 1.95($\Delta$) & 4.8 & 3/2 & odd \\
& & & & & \\
\hline
& & & & & \\
$m+n+k=4$ & 2.17 & 2.19($N$) & 0.9 & 1/2 & even \\
$m+n+k=4$ & 2.17 & 2.22($N$) & 2.3 & 1/2 & even \\
& & & & & \\
$m+n+k=5$ & 2.48 & 2.42($\Delta$) & 2.4 & 3/2 & odd \\
& & & & & \\
\hline
& & & & & \\
$m+n+k=6$ & 2.79 & 2.65($N$) & 5.0 & 1/2 & even \\
$m+n+k=6$ & 2.79 & 2.85($\Delta$) & 2.2 & 3/2 & even \\
& & & & & \\
\hline
& & & & & \\
$m+n+k=7$ & 3.10 & 3.03($N$) & 2.3 & 1/2 & odd \\
$m+n+k=7$ & 3.10 & 3.23($\Delta$) & 4.2 & 3/2 & odd \\
& & & & & \\
... & ... & ...& ... & & \\
\hline\hline
\end{tabular}
\end{center}
\vskip .6in
\begin{center}
\begin{verse}
Table 7. Baryon states $N$ and $\Delta$. The energies $E_{C}$ were \\
calculated according to the formula $E_{n,m,k} = 0.31(n+m+k+3)$.\\
$E_{M}$ is the measured energy. A few states were not reproduced \\
with this simple treatment. The error means the absolute value of \\
$(E_{C} - E_{M})/E_{C}$.\\
\end{verse}
\end{center}

\newpage
\begin{center}
\begin{tabular}{ c  l l l c c } \hline
& & & & & \\
State($m,n,k$) & $E_{C}(Gev)$ & $E_{M}$(Gev) & Error{\%} & Isospin & Parity \\
& & & & & \\
\hline\hline
& & & & & \\
0,0,0  & 1.12  & 1.1156($\Lambda$) & 0.4 & 0 & even\\
0,0,0  & 1.12  & 1.1894($\Sigma$) & 6.2 & 1 & even \\
0,0,0  & 1.12  & 1.1925($\Sigma$) & 6.5 & 1 & even \\
0,0,0  & 1.12  & 1.1973($\Sigma$) & 6.9 & 1 & even \\
\hline
$m+n=1$, k=0 & 1.43 & 1.385($\Sigma$) & 3.2 & 1 & odd \\
$m+n=1$, k=0 & 1.43 & 1.405($\Lambda$) & 1.7 & 0 & odd \\
\hline
0,0,1 & 1.62 & 1.67($\Sigma$) & 3.1 & 1 & odd \\
0,0,1 & 1.62 & 1.67($\Lambda$) & 3.1 & 0 & odd \\
\hline
$m+n=2$, k=0 & 1.74 & 1.69($\Lambda$) & 2.9 & 0 & even \\
$m+n=2$, k=0 & 1.74 & 1.75($\Sigma$) & 0.6 & 1 & even \\
$m+n=2$, k=0 & 1.74 & 1.765($\Sigma$) & 1.4 & 1 & even \\
\hline
$m+n=1$, k=1 & 1.93 & 1.915($\Sigma$) & 0.8 & 1 & even \\
$m+n=1$, k=1 & 1.93 & 1.94($\Sigma$) & 0.5 & 1 & even \\
\hline
$m+n=3$, k=0 & 2.05 & 2.03($\Sigma$) & 1.0 & 1 & odd \\
\hline
0,0,2 & 2.12 & 2.1($\Lambda$) & 0.9 & 0 & even \\
\hline
$m+n=2$, k=1 & 2.24 & 2.25($\Sigma$) & 0.5 & 1 & odd \\
\hline
$m+n=4$, k=0 & 2.36 & 2.35($\Lambda$) & 0.4 & 0 & even \\
\hline
$m+n=1$, k=2 & 2.43 & 2.455($\Lambda$) & 1.0 & 0 & odd \\
\hline
0,0,3 & 2.62 & 2.585($\Lambda$) & 1.3 & 0 & odd \\
0,0,3 & 2.62 & 2.62($\Sigma$) & 0 & 1 & odd \\
\hline
... & ... & ... & ... & &  \\
\hline\hline
\end{tabular}
\end{center}
\vskip .6in

\begin{center}
\begin{verse}
Table 8. Baryon states $\Sigma$ and $\Lambda$. The energies $E_{C}$ were\\
calculated according to the formula $E_{n,m,k}= 0.31(n+m+2) + 0.5(k+1)$.\\
$E_{M}$ is the measured energy. A few states were not reproduced by this \\
simple treatment. The error  means the absolute value of $(E_{C} -
E_{M})/E_{C}$. \\
\end{verse}
\end{center}

\newpage

\begin{center}
\begin{tabular}{ c  l l l c c }
\hline
& & & & & \\
State($m,n,k$) & $E_{C}$(Gev) & $E_{M}$(Gev) & Error(\%) & Isospin & Parity \\
& & & & & \\
\hline
\hline
& & & & & \\
0,0,0 & 1.31 & 1.315 & 0.4 & 1/2 & even \\
0,0,0 & 1.31 & 1.321 & 0.8 & 1/2 & even \\
& & & & & \\
\hline
& & & & & \\
1,0,0 & 1.62 & 1.53 & 5.6 & 1/2 & odd \\
& & & & & \\
\hline
& & & & & \\
$n=0,m+k=1$ & 1.81 & 1.82 & 0.6 & 1/2 & odd \\
& & & & & \\
\hline
& & & & & \\
2,0,0 & 1.93 & 1.94 & 0.5 & 1/2 & even \\
& & & & & \\
\hline
... & ... & ... & ... & ... & ... \\
\hline\hline
\end{tabular}
\end{center}
\vskip .6in

\begin{center}
\begin{verse}
Table 9. Baryon states $\Xi$. The energies $E_{C}$ were calculated  \\
according to the formula $E_{n,m,k} = 0.3(n+1) + 0.5(m+k+2)$.  $E_{M}$ \\
is the measured energy. The error is the absolute value of
$(E_{M} - E_{C})/E_{C}$.
\end{verse}
\end{center}
\newpage
\noindent
Complete address:
\vskip .3in
\noindent
M\'{a}rio Everaldo de Souza,
\newline
\noindent
Universidade Federal de Sergipe
\newline
\noindent
Departamento de F\'{\i}sica - CCET,
\newline
\noindent
49000 Aracaju, Sergipe, Brazil
\newline
\noindent
Phone no. (55)(79)241-2848, extension 354
\newline
\noindent
Fax no. (55)(79)241-3995
\newline
\noindent
e-mail DFIMES@BRUFSE.BITNET
\end{document}